\documentclass[11pt]{article}

\usepackage{amsfonts}
\usepackage{amssymb}
\usepackage{amsmath}
\usepackage{amsthm}
\usepackage{latexsym}
\usepackage{verbatim}
\usepackage{epsfig}
\usepackage{amsbsy}
\usepackage{amscd}
\usepackage{bm}

\usepackage{graphicx}

\usepackage{ifthen} 
\usepackage{xcolor}
\usepackage[colorlinks=true, linkcolor=teal, anchorcolor=teal,
            citecolor=blue,  filecolor=blue, menucolor=blue, pagecolor=teal,											urlcolor=blue, plainpages=false, pdfpagelabels]{hyperref}

\newcommand{\mymail}[1]{\href{mailto:#1}{\texttt{#1}}}

\usepackage{fancyhdr}
\usepackage[realmainfile]{currfile}
\fancypagestyle{firststyle}{

\cfoot{{\footnotesize $^\ast$: Partially supported by NSF grants \#2120318 and \#2411270}} 
}

\newcommand{\setauthA}[1]{\def\authA{#1}}

\def\printA{\begin{tabular}{l} \authA \end{tabular}}

\newcommand{\makemytitle}[1]{\begin{center}{\textsf{\LARGE #1}}
  \end{center}
}

\usepackage[sf,bf]{titlesec}

\usepackage{algorithmic}
\usepackage{algorithm}

\usepackage[nonamebreak,square,numbers,sort]{natbib}

\usepackage[letterpaper, textheight = 676pt]{geometry}
\providecommand{\M}[1]{\mathbf#1}
\providecommand{\mc}[1]{\mathcal#1}
\providecommand{\mc}[1]{\mathcal#1}

\newcommand{\R}{{\mathbb R}}

\DeclareMathOperator{\E}{\mathbf{E}}
\DeclareMathOperator{\p}{\mathbf{P}}

\DeclareMathOperator{\cov}{Cov}

\newcommand{\indep}{\rotatebox[origin=c]{90}{$\models$}}

\providecommand{\T}{\top} 

\DeclareMathOperator*{\argmin}{argmin}

\providecommand{\wt}[1]{\widetilde{#1}}
\providecommand{\wh}[1]{\widehat{#1}}

\newcommand{\blanco}[1]{  }

\newcommand{\deriv}[3]{%
\ifthenelse{#1 = 1}{\frac{d\,#2}{d\,#3}}{\frac{d^{{#1}} #2}{d{#3}^{{#1}}}}
}

\newcommand{\partials}[3]{%
\ifthenelse{#1 = 1}{\frac{\partial\,#2}{\partial\,#3}}{\frac{\partial^{#1}
    #2}{\partial#3^{#1}}}
} 

\def\su{\sum_{i=1}^n}

\def \coloneq{\mathrel{\mathop:}=}
\def \invcoloneq{=\mathrel{\mathop:}}
\def \eps{\varepsilon}

\newtheorem{propo}{Theorem}

\newtheorem{prop}[propo]{Proposition}

\def\R{\mathbb{R}}

\def\eps{\epsilon}

\usepackage{booktabs}
\usepackage{multirow}

\usepackage{subfigure}
\usepackage{caption}
\newboolean{isjournal}
\setboolean{isjournal}{true}

\newboolean{nocolors}
\setboolean{nocolors}{false}

\newboolean{usemtpro}
\setboolean{usemtpro}{false}

\ifthenelse{\boolean{nocolors}}{\hypersetup{colorlinks=false}}{}

\makeatletter
\newcommand\footnoteref[1]{\protected@xdef\@thefnmark{\ref{#1}}\@footnotemark}
\makeatother

\setauthA{{\bfseries Martin Slawski}$^{*}$}

\begin{document}
\thispagestyle{firststyle}

\makemytitle{\begin{tabular}{c}{{\Large {\bfseries Causal Secondary Analysis of Linked Data }}} \\ {\Large {\bfseries in the Presence of Mismatch Error}}\end{tabular}}
\vskip 3.5ex
%
{\large\begin{center}
\printA
\vskip.5ex
{\scriptsize Department of Statistics, University of Virginia, Charlottesville, VA 22903, USA $\; \; \;$}
{\small \mymail{ebh3ep@virginia.edu}} 
\end{center}}
\vskip 3.5ex

\begin{abstract} 
The increased prevalence of observational data and the need to integrate 
information from multiple sources are critical challenges in contemporary
data analysis. Record linkage is a widely used tool for combining datasets in the absence of unique identifiers. The presence of linkage errors such
as mismatched records, however, often hampers the analysis of data sets
obtained in this way. This issue is more difficult to address in secondary analysis settings, where linkage and subsequent analysis are performed separately, and analysts have limited information about linkage quality. In this paper, we investigate the estimation of average treatment effects in the conventional potential outcome-based causal inference framework under linkage uncertainty. To mitigate the bias that would be incurred with naive
analyses, we propose an approach based on estimating equations that treats the unknown match status indicators as missing data. Leveraging a variant of the Expectation-Maximization algorithm, these indicators are imputed 
based on a corresponding two-component mixture model. The approach is amenable to asymptotic inference. Simulation studies and a case study highlight  
the importance of accounting for linkage uncertainty and demonstrate the effectiveness of the proposed approach. 
\end{abstract}
\vspace*{-.5ex}
\vspace*{-1ex}
\section{Introduction}\label{sec:intro}
\vspace*{-1ex}
The digital revolution has not only led to an explosion in data volumes but has also profoundly transformed the ways in which data are collected and analyzed. One fundamental change concerns the transition from strict experimental design and sampling protocols towards the use of already existing data. The analysis
of such observational data sets tends to be more challenging since in most cases
specific steps need to be taken to avoid biases that would normally be addressed at the design and data collection stage. Causal inference \cite[e.g.,][]{Imbens2015, Rosenbaum2010, Hernan2010book} has emerged as the field 
dedicated to address this challenge. 

Another trend related to data collection concerns the integration of data from multiple sources with the goal of creating richer data sets that provide a more 
comprehensive view on questions of interest. Record Linkage (RL, \cite[e.g.,][]{Newcombe, Binette2022, Christen2012}) is a key technique for this task, enabling the record-by-record combination of two data sets pertaining to a common set 
of statistical units in the absence of exact identifiers. In past years, RL has 
seen widespread use across disciplines. Examples include the linkage of surveys and 
administrative records \cite{abowd2021, Enamorado2019}, electronic health records and insurance billing information \cite{N3C}, criminal justice data \cite{DataFirst}, and historical vital records and censuses \cite{LifeMcit}. 

The subject of this paper concerns the intersection of causal inference and RL. Specifically, we study the estimation of average treatment effects (ATEs) based on classical estimators in the presence of incorrect links in the linked data set under consideration. Such mismatched records, mismatches for short, are common when the set of quasi-identifiers (also known as ``matching variables") used to identify pairs of records belonging to the same entity are ambiguous or prone to errors. For instance, regulations such as the Health Insurance Portability and Accountability Act (HIPAA) typically limit the amount of personally identifying information to ZIP Code, Date of Birth, and Sex. Matching on names or addresses is generally not reliable due to recording errors, changes of this information over time, or the commonness of certain person or street names. As well-documented in the literature \cite{Neter65, Lahiri05, Wang2022, Chambers2023, Kamat2024}, ignoring mismatches can adversely effect downstream statistical analysis and may lead to invalid findings. Performing suitable adjustments is arguably more difficult in the setting of {\em secondary analysis} in which linkage and subsequent analysis are performed separately, and only the linked file with (at best) limited information about the quality of each link is available to the data analyst. The secondary analysis setting has become more pervasive in recent years \cite{Slawski2024}, as a result 
of stronger incentives to share research data and increased tendencies to involve expert third-party services for performing linkages. This in turn prompts the development of statistical tools for {\em post-linkage} (i.e., downstream) analysis 
accounting for uncertainty and errors at the linkage stage. 
\vskip1.5ex
\noindent {\bfseries Contributions and Related Work}. There is a growing body
of literature on post-linkage analysis as summarized in the recent review paper \cite{Kamat2024}; the bulk of this literature concerns regression analysis with predictor variables in one and the response variable in another file. Concerning the secondary analysis setting, the paper \cite{Kamat2024} roughly distinguishes two lines of work: (i) weighting 
methods and (ii) likelihood or Bayesian methods revolving around a two-component mixture model capturing correctly and incorrectly linked records, respectively. In a nutshell, the prevalent variant \cite{Chambers2009} of approach (i) is based on the construction of unbiased estimating equations in which the original predictors and responses are replaced by weighted combinations thereof; the weights are obtained under the so-called exchangeable linkage error (ELE) assumption, which postulates that mismatches occur uniformly at random within blocks defined by matching variables required
to agree exactly for any linked pair of records. A shortcoming of (i) is that 
estimates of block-wise linkage error rates need to be available. Roughly two variants can be delineated when considering approach (ii): one variant entails
the explicit specification of models for each of the two components \cite{gutmanmixture}, whereas in the 
other variant the component associated with mismatches is fixed by assuming independence of (the incorrectly linked) predictors and responses \cite{Slawski2024}. 
Compared to (i), approach (ii) is more flexible and efficient, but also more sensitive to model misspecification.  

Prior works studying post-linkage causal analysis concern the primary
analysis setting in which linkage and downstream analysis are performed jointly. In \cite{Shan2021}, the authors consider linkage of two files with
one file containing covariates and outcomes while the second file contains 
a list of individuals that received the (binary) treatment and
additional covariates observed for those individuals. This setting is related
yet different from our Scenario III described below. A Bayesian framework is
adopted in \cite{Shan2021} in which the uncertain treatment status is multiply imputed before using a combination of propensity score matching and model-based imputation of potential outcomes for the causal analysis. In \cite{Guha2022}, the author consider our Scenario I in which treatment status 
and covariates are contained in one data set while the outcomes are contained
in a separate data set. A Bayesian joint model encompassing probabilistic record linkage and regression on propensity scores for imputing potential outcomes is proposed. In follow-up work, \cite{Guha2024} study the case in which some of the covariates are contained in a separate file. A combination of propensity score and regression-based estimators are applied to infer treatment effects, alongside Markov Chain Monte Carlo-based multiple imputation of the match status for pairs of records to be linked.   

In the present paper, we consider similar setups in the {\em secondary} analysis setting. The three quantities of interest (outcomes, exposure/treatment, covariates) are assumed to be spread over two files, which 
prompts three scenarios to be studied (cf.~$\S$\ref{subsec:setup}). Only the linked file, which may include auxiliary covariates informing the match status of the linked records, is available. We quantify the bias of certain naive
propensity-score based estimators, namely (i) the estimator that uses only records known to be correctly matched as well as (ii) the estimator ignoring 
incorrect links. We develop a framework based on estimating equations in which the unknown record-wise match status indicators (i.e., correctly vs.~incorrectly linked) appear as missing variables. This framework yields 
propensity score and regression-based estimators of ATEs that adjust for the presence of mismatched records and also enables asymptotic inference. The imputation of the missing variables relies on a variant of the Expectation-Maximization (EM) algorithm for estimating equations in conjunction with a two-component mixture model akin to the approach in \cite{Slawski2024}. Compared to the latter work, we herein relax the assumption of strongly informative mismatch error \cite{Kamat2024, Bukke2025}. We also study the impact of different types of model misspecification and associated remediation strategies. Despite the apparent differences in setups, the approach in the present paper exhibits significant high-level conceptual overlap with \cite{Guha2024}. In simplified terms, both studies adjust conventional ATE
estimators for linkage errors using different approaches for imputing match status depending on the available information and the chosen inferential framework.  
\vskip1ex
\noindent {\bfseries Organization}. Our methodology is presented in 
$\S$\ref{sec:meth}, which is divided into multiple sections. $\S$\ref{subsec:setup} introduces our setup and the associated
assumptions. $\S$\ref{subsec:PSest} is dedicated to propensity score
estimators in the setup under consideration, including a quantification of the bias of two naive estimators that discard and ignore uncertain links, respectively. $\S$\ref{subsec:model-DR} presents the heart of our inferential
framework based on estimating equations with latent variables. Model misspecification is discussed in $\S$\ref{subsec:modelmis}. Simulation studies 
are contained in $\S$\ref{sec:simulation}. A case study with real data is
presented in $\S$\ref{sec:casestudy}. We conclude in $\S$\ref{sec:conclusion}. Proofs and technical details can be found in the appendix.

\vskip1.5ex
\noindent {\bfseries Notation}. For the convenience of the reader, a summary of frequently used notation is tabulated below.  
\begin{table}[h!!!]
{\footnotesize
\begin{center}
\begin{tabular}{|ll|ll|}
\hline & & &\\[-1.5ex]
$E$ & (binary) exposure/treatment   & $e_i$ & exposure for obs.~$i$ \\[.5ex]
$Y(e)$ & potential outcome if $E = e$  & $y_i$ & outcome for obs.~$i$ \\[.5ex]
$Y$    & observed outcome & & \\
$X$    & covariates (causal) & $\M{x}_i$ & covariates (causal) for obs.~$i$  \\[.5ex]
$Z$    & covariates (linkage) & $\M{z}_i$ & covariates (linkage) for obs.~$i$  \\[.5ex]
$M$ & Mismatch Indicator  & $m_i$ &  mismatch indicator for obs.~$i$ \\[.5ex]
$\mu_{\bm{\beta}}^1(\M{x})$ & outcome model for $Y(1)$  & $\bm{\beta}$ & parameter for outcome model(s)  \\[.5ex]
$\mu_{\bm{\beta}}^0(\M{x})$ & outcome model for $Y(0)$  & &  \\[.5ex]
$p_{\bm{\phi}}(\M{x})$ & propensity score (PS) model  & $\bm{\phi}$  & parameter for PS model \\[.5ex]
$h_{\bm{\gamma}}(\M{z})$ & model for $M$  & $\bm{\gamma}$ & parameter for the $M$-model  \\[.5ex]
$\varphi_{\sigma}$ & PDF of the $N(0,\sigma^2)$ dist. & $\tau$ & average treatment effect \\
\hline
\end{tabular}
\end{center}}
\vspace*{-3ex}
\caption{Summary of notation used repeatedly in this paper.}\label{tab:notation}
\end{table}\\[-2ex]
We adopt the potential outcomes framework and the associated notation in \cite{Imbens2015}, with 
$Y(e)$ denoting the outcome that would be observed under
exposure (or treatment) status $E = e$, $e = 0,1$. The observed outcome $Y$ equals the potential outcome according to the realized exposure status. We use uppercase letters such 
as $E$, $Y$, etc.~to refer to the underlying random variables, while lowercase (and potentially boldfaced) letters are used for fixed values in the range of these
random variables as well as for the observed data. For the latter, we do not 
use separate notation to distinguish random variables and their realizations since the distinction can be inferred from the context.  

\section{Methodology}\label{sec:meth}

\subsection{Setup}\label{subsec:setup}
We start by pinning down our setup(s) and basic assumptions. First, we assume consistency, i.e., 
$E = e$ implies that $Y = Y(e)$ is observed, and no interference between different statistical units, i.e., the stable unit treat value assumption (SUTVA, \cite[][$\S$1.6.1]{Imbens2015}) holds true. The following three scenarios of record linkage are considered in the sequel. These scenarios are not comprehensive: e.g., one might also consider a separation into three files, subsets of covariates spread across the two files as in \cite{Guha2024}, or the second file representing the subset of treated units as in \cite{Shan2021}. At the same time, the three scenarios to be studied represent a natural starting point, with one of the three integral pieces of information contained in a separate file in each case. \\
\begin{equation*}
\begin{array}{|ll|} \hline
\textsc{Scenario I} & \\[1ex]
\text{{\bfseries File A}} & \text{{\bfseries File B}} \\
\boxed{\M{x} \;\; e}            & y \\[1ex] \hline
\end{array} \qquad \quad
\begin{array}{|ll|} \hline
\textsc{Scenario II} & \\[1ex]
\text{{\bfseries File A}} & \text{{\bfseries File B}} \\
\M{x}            & \boxed{y \; \; e} \\[1ex]
\hline
\end{array} \qquad \quad \begin{array}{|ll|} \hline
\textsc{Scenario III} & \\[1ex]
\text{{\bfseries File A}} & \text{{\bfseries File B}} \\
\boxed{\M{x} \; \; y}            & e \\[1ex]
\hline
\end{array}
\end{equation*}
\vskip2ex
\noindent We suppose that record linkage is used to merge the two files, yielding
a single linked file consisting of triplets $\{ (\M{x}_i, e_i, y_i) \}_{i = 1}^n$ to be used 
for analysis. Linkage is generally not error-free, i.e., some of the triplets may result
from an incorrect pairing of records (mismatches) across the two files that do not correspond to the same statistical unit, in which case the associated (latent) mismatch indicators $\{ m_i \}$ take the value one (and zero otherwise). In the setting of secondary analysis, linkage is considered a ``black box"; only its output, i.e., the linked file and possibly covariates $\{ \M{z}_i \}_{i = 1}^n$ informative of the match status $\{ m_i \}_{i = 1}^n$ are available to the analyst. These linkage-related covariates may be observed jointly with the 
aforementioned $\{ (\M{x}_i, e_i, y_i) \}$-triplets, and are supposed to be unrelated to the outcome, as formalized in ({\bfseries A3}) below. 

\begin{figure}
\begin{center}
\includegraphics[width = .4\textwidth]{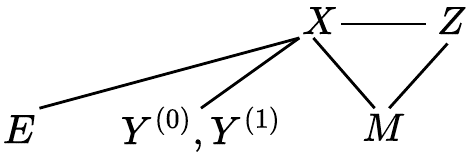}
\end{center}
\vspace{-2ex}
\caption{Diagram capturing the association structure of the variables of interest.}\label{fig:causal_diagram}
\end{figure}

\begin{itemize}
\item[({\bfseries A1})] {\em Unconfoundedness}. $\{Y(1), Y(0) \} \indep \{M, E\} | X$,
\item[({\bfseries A2})] $M \indep \, E | X$,
\item[({\bfseries A3})] $\{ Y(1), Y(0), E\} \indep \, Z | X$.
\item[({\bfseries A4})] $0 < \p(E = 1 | X = \M{x}) < 1$ for all $\M{x}$.
\item[({\bfseries A5})] Given any {\em incorrectly} linked record of the form $(\M{a}, \M{b})$ with variables $\M{a}$ and $\M{b}$ originating in Files A and B, respectively, then $\M{a} \indep \M{b}$.  
\end{itemize}
A diagram incorporating {\bfseries (A1)} through {\bfseries (A3)} is provided in Figure \ref{fig:causal_diagram}.
\vskip1ex
\noindent {\bfseries Discussion}. Note that ({\bfseries A1}) is akin to the usual 
``no unobserved confounding" assumption in the literature \cite[][Defn.~3.6]{Imbens2015}, which reads $\{Y(1), Y(0) \} \indep E | X$. We here add $M$ on the right hand side of the independence symbol to ensure that conditional on $X$ the potential outcomes are unaffected by the linkage process. Without the latter assumption, it is unclear how to identify 
the causal estimand of interest $\tau^* = \E[Y(1)] - \E[Y(0)]$, subsequently referred to as the average treatment effect (ATE). To see this, consider the regression models 
$Y(1) = \alpha_1 + \beta_1 X + b \cdot M + \eps$ and $Y(0) = \alpha_0 + \beta_0 X + \eps$ with $\eps$ representing unstructured noise. Note that inferring the parameters of the first regression model entails the restoration of the correct pairings of the mismatched records, which is generally not feasible \cite{Pananjady2017, Slawski2019}. 


By contrast, assumption {\bfseries (A2)} is made for convenience, yielding a simple product form for the propensity 
score estimators in $\S$\ref{subsec:PSest} consisting of a conventional propensity score and a mismatch probability. As indicated above, {\bfseries (A3)} singles out a subset of covariates related to $M$ only. Note that the latter may still depend on $X$ as well. 

Assumption {\bfseries (A4)} is a standard regularity assumption on the propensity score \cite{Rosenbaum1983} 
$\p(E = 1 | X = \M{x}) \invcoloneq p_{\bm{\phi}}(\M{x})$, asserting that for any configuration 
of the covariates, there is a non-zero probability of (not) being in the exposure group \cite[cf.][$\S$12.2]{Imbens2015}. 

Finally, in the RL literature {\bfseries (A5)} is proposed in \cite{Slawski2021, Slawski2024}. It states that the two pieces $\M{a}$ and $\M{b}$ obtained through file linkage are independent if the associated link is incorrect, i.e., the two pieces actually belong
to different statistical units. This assumption is automatically satisfied in the situation that the fragments contained in the two files originate from independent triplets $(\M{x}, e, y)$ whose three components belong to the same statistical unit. While generally plausible, there are settings in which {\bfseries (A5)} may not hold, e.g., if linked pairs are required to agree
on certain covariates.  

\vskip2ex
\noindent {\em Audit Sample.} While in general, the match status $m_i$ of observation $i$ is unknown, $1 \leq i \leq n$, there might be a subsample of observations $\mc{A} \subseteq \{1,\ldots,n\}$ for which the $\{ m_i \}$ are observed. We refer to $\mc{A}$ as ``audit sample" that could have been obtained from an ``audit" of linked records by a reviewer verifying the correctness of the matches, potentially with access to additional (external) information. The following two simplifying assumptions are made in this regard (i) membership in $\mc{A}$ is 
supposed to be independent of all variables under information, and (ii) the reviewer does not mislabel 
any of the records, i.e., the reviewer's assessment always coincides with the true values $\{ m_i \}_{i \in \mc{A}}$. As will become more clear in $\S$\ref{subsec:modelmis} below, the potential usefulness of the audit sample lies in the fact that consistent estimation of the ATE is no longer contingent on having correctly specified models for the potential outcomes.

\subsection{Propensity Score Estimators}\label{subsec:PSest}

\begin{figure}
\includegraphics[width= 0.24\textwidth]{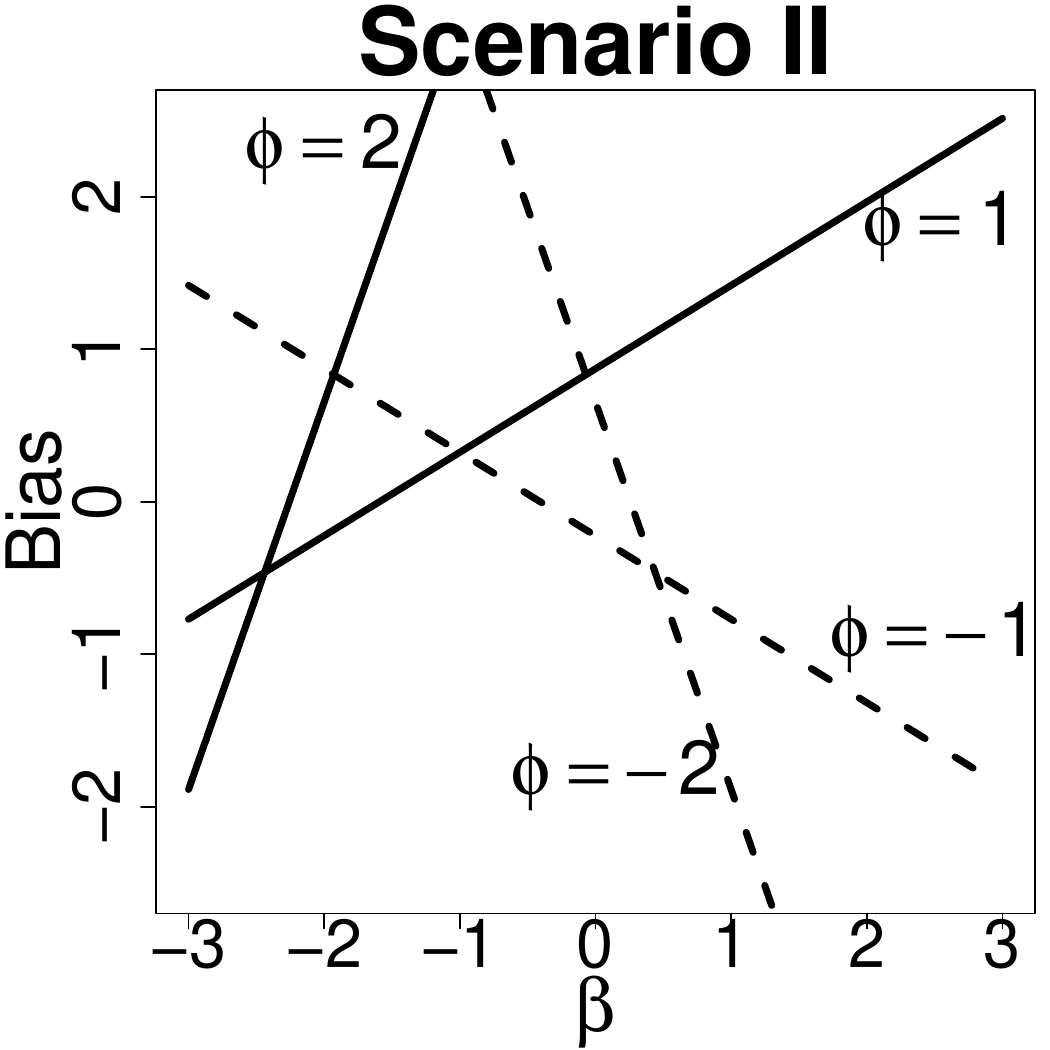}  
\includegraphics[width= 0.24\textwidth]{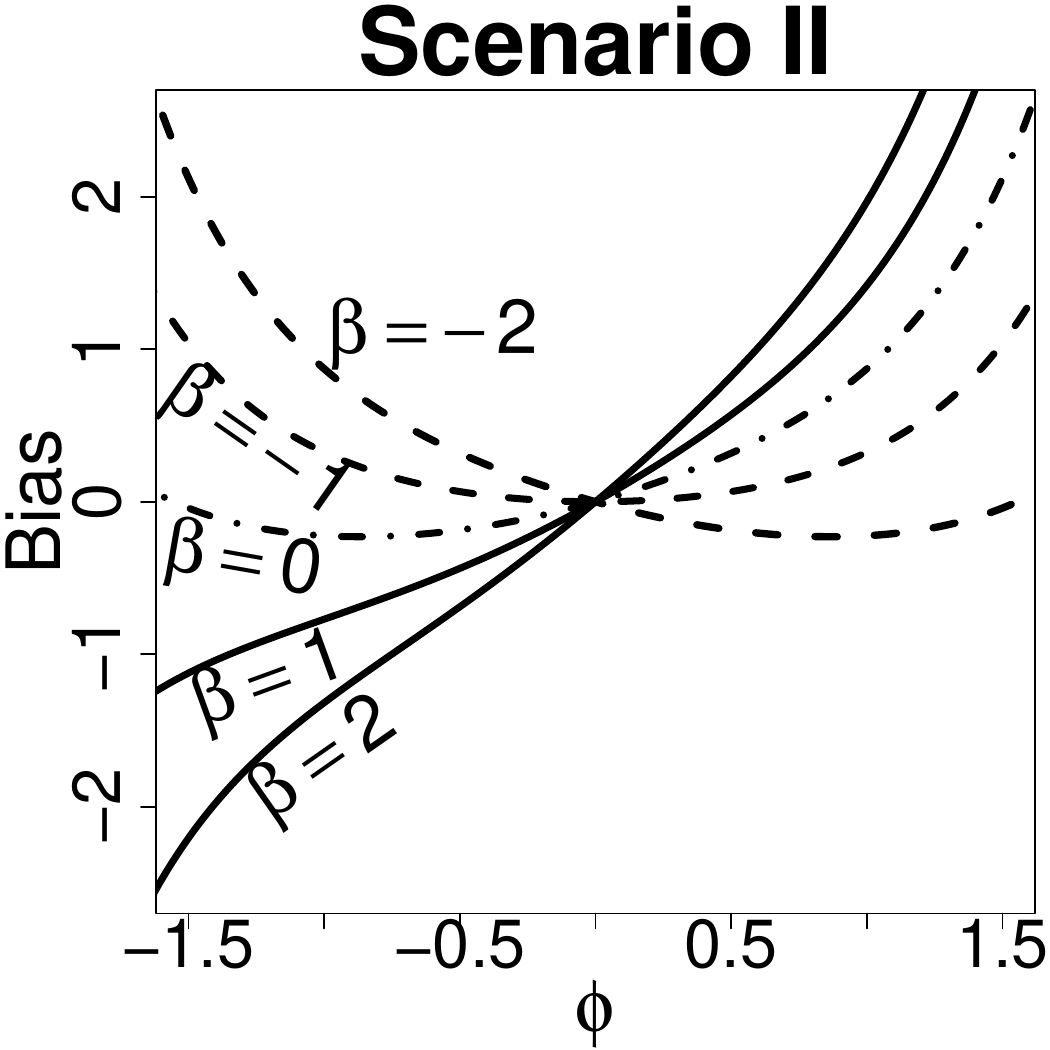} 
\includegraphics[width= 0.24\textwidth]{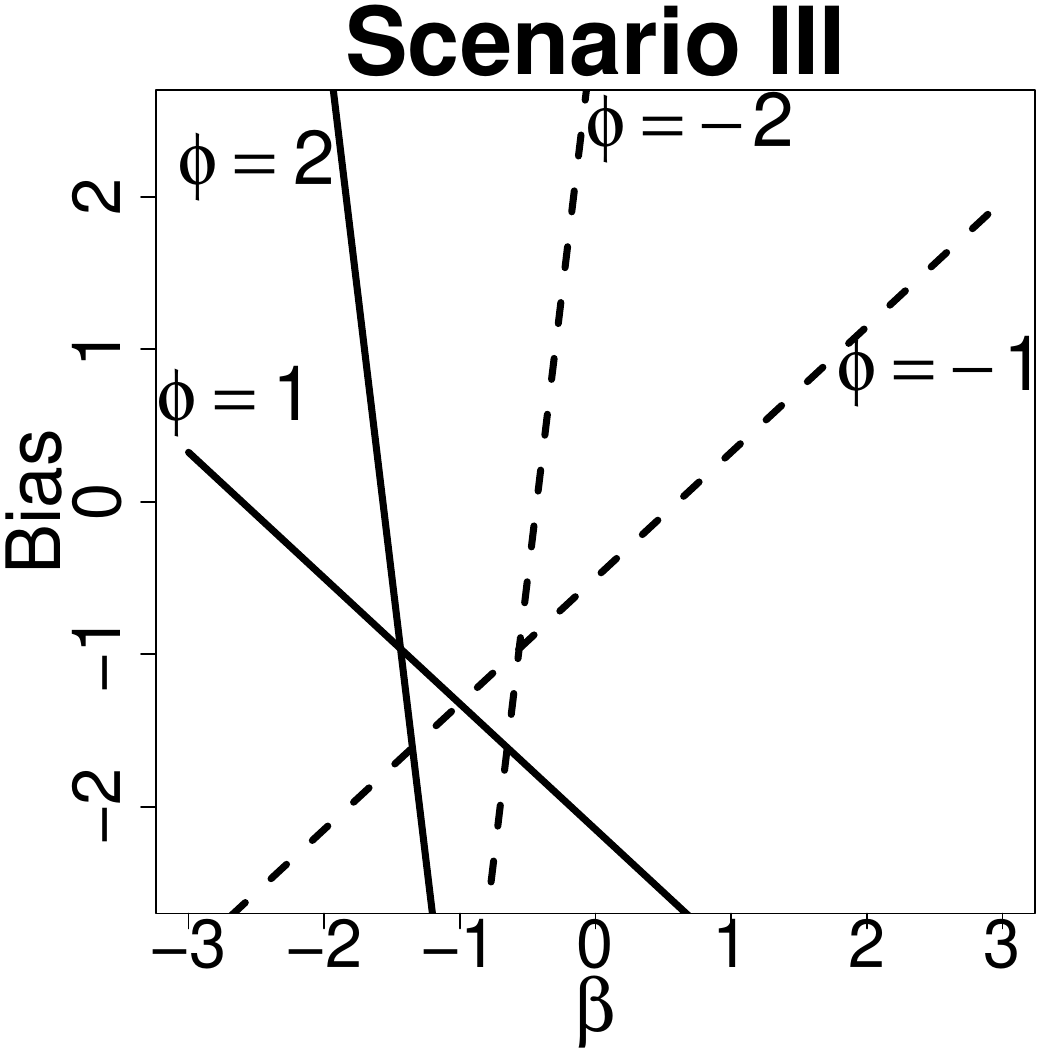} 
\includegraphics[width= 0.24\textwidth]{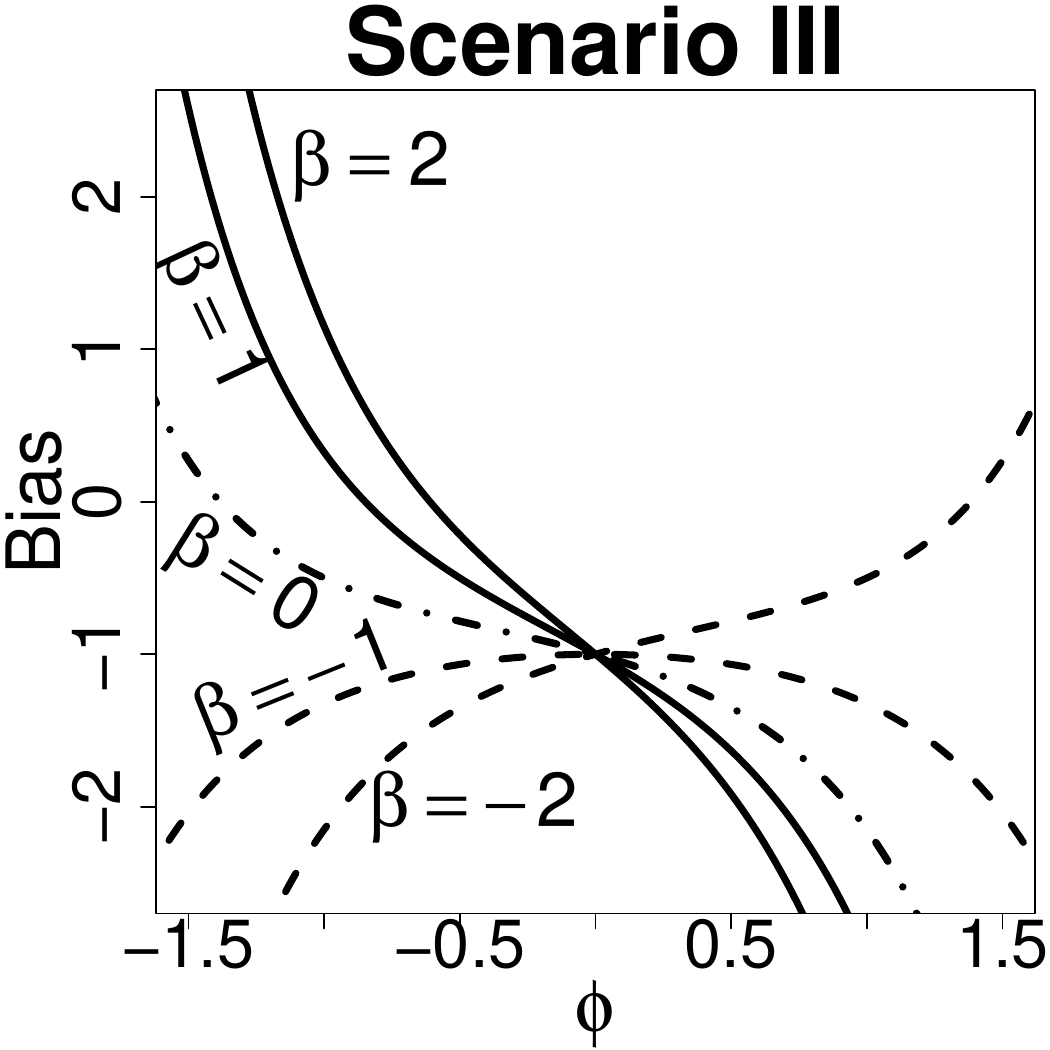}
\vspace*{-1ex}
\caption{Bias (divided by the mismatch rate $\alpha$) in estimating the ATE $\tau^*$ according to Proposition \ref{prop:bias_naive} as a function of $\beta$ and $\phi$ when $\mu_0(x) = \E[Y^{(0)}|X = x] = x$, $\mu_1(x) = \E[Y^{(1)}|X=x] = \beta x + 1$, $p_{\phi}(x) = \p(E = e|X = x) = 1/\{1 + \exp(-x \phi)\}$, $X \sim N(0,1)$.}\label{fig:bias}
\end{figure}

{\em Naive estimation I}. We start by assuming for simplicity that the underlying propensity score model $p_{\bm{\phi}}$ is known and
correctly specified. Even in this case, ignoring mismatch error when constructing the usual PS estimators of the ATE will typically yield biased estimates. Consider the simple situation in which the match status $M$ is independent of all other variables with $\p(M = 1) = \alpha$. The standard PS estimator is given 
by 
\begin{equation*}
\wh{\tau}_{\text{naive}} = \left( \sum_{i: e_i = 1} \frac{y_i}{p_{\bm{\phi}}(\M{x}_i)} -  \sum_{i: e_i = 0} \frac{y_i}{1 - p_{\bm{\phi}}(\M{x}_i)} \right) \Big / n. 
\end{equation*}
The following statement quantifies the bias of $\wh{\tau}_{\text{naive}}$, separately for each scenario. 
\begin{prop}\label{prop:bias_naive} In addition to Assumptions {\em ({\bfseries A1})} and {\em({\bfseries A3})--({\bfseries A5})}, suppose that $M \sim \text{{\em Bernoulli}}(\alpha)$. We then have \\
(i) Under Scenario I ($\M{x}$ and $e$ in the same file):
\begin{align*}
\E[\wh{\tau}_{{\text{{\em naive}}}}] = (1-\alpha) \tau^*. 
\end{align*}
(ii) 
Under Scenario II ($e$ and $y$ in the same file):
\begin{align*}
\E[\wh{\tau}_{{\text{{\em naive}}}}] = (1-\alpha) \tau^* + \alpha \left( \E_{X,X'} \left[\mu_1(X) \frac{p_{\bm{\phi}}(X)}{p_{\bm{\phi}}(X')} \right]  - \E_{X,X'} \left[\mu_0(X) \frac{1 - p_{\bm{\phi}}(X)}{1 - p_{\bm{\phi}}(X')} \right]  \right),
\end{align*}
where $X'$ is independent of and identically distributed as $X$, and $\mu_e(X) = \E[Y^{(e)}|X]$, $e = 0,1$.\\ 
(iii) Under Scenario III ($\M{x}$ and $y$ in the same file):
\begin{align*}
\E[\wh{\tau}_{{\text{{\em naive}}}}] &= (1-\alpha) \tau^* + \alpha p \E[Y^{(1)}] - \alpha (1-p) \E[Y^{(0)}] + \\
&\quad + \alpha p \E_X \left[\mu_0(X) \frac{1 - p_{\bm{\phi}}(X)}{p_{{\bm{\phi}}}(X)} \right]  - \alpha (1-p) \E_X \left[\mu_1(X) \frac{p_{\bm{\phi}}(X)}{1 - p_{{\bm{\phi}}}(X)} \right],
\end{align*}
where $p = \p(E = 1)$. 
\end{prop}
\noindent Note that in the special case $p_{\bm{\phi}}(\M{x}) \equiv 0.5$, the expression in (iii) simplifies to $(1 - \alpha) \tau^*$ as in (i), i.e., the ATE estimates undergoes attenuation, which is a well-known consequence of mismatch error \cite{Neter65, Wang2022}. To an extent, Scenario II appears the ``most benign": e.g., the bias is zero whenever $p_{\bm{\phi}}$ is constant. In general, the expressions in (ii) and (iii) depend heavily on the distribution of $X$ and the form of the $\mu$s and $p_{\bm{\phi}}$. In Figure \ref{fig:bias}, we evaluate the bias for Gaussian $X$ and linear and logistic models for the $\mu$s and $p_{\bm{\phi}}$, respectively. The figures show that even small fractions of mismatches may result in substantial bias.    
\vskip1ex
\noindent {\em Naive estimation II}. Since the match indicators for the audit sample $\mc{A}$ are known, it is tempting to use exclusively the subset of correct matches $\mc{A}_{0} = \{i \in \mc{A}: \; m_i = 0\}$ for PS estimation, prompting the estimator  
\begin{equation*}
\wh{\tau}_{\mc{A}_0} = \Bigg( \sum_{\substack{i \in \mc{A}_0 \\ e_i = 1}} \frac{y_i}{p_{\bm{\phi}}(\M{x}_i)}  - \sum_{\substack{i \in \mc{A}_0 \\ e_i = 0}} \frac{y_i}{1 - p_{\bm{\phi}}(\M{x}_i)} \Bigg) \Big/|\mc{A}_0|.  
\end{equation*}
However, this estimator generally incurs a bias as well since $M$ might be associated 
with $X$ and hence may constitute a selection mechanism in the same way as $E$ does. In the simplest 
case with a completely randomized treatment assignment and thus $E$ independent of $X$ (and $M$) and $Z = X$, 
we have
\begin{equation*}
\E[Y I(M = 0) I(E=e)] = \E[ \E[Y^{(e)}|X] (1-h_{\bm{\gamma}}(X))] \p(E = e), \; e \in \{0,1\},
\end{equation*}
which is not proportional to $\E[Y^{(e)}]$ in general. For example, if $X$ is $\{0,1\}$-valued then the first term becomes $$\E[Y^{(e)} | X = 1] (1 - h_{\bm{\gamma}}(1)) \p(X = 1)+ \E[Y^{(e)} | X = 0] (1 - h_{\bm{\gamma}}(0)) \p(X = 0), \; e \in \{0,1\},$$ which yields a weighted average of $\E[Y^{(e)} | X = 1]$ and $\E[Y^{(e)} | X = 0]$ so that if (w.l.o.g) $h_{\bm{\gamma}}(1) > h_{\bm{\gamma}}(0)$, one observes a bias towards $\E[Y^{(e)}|X = 0]$, i.e., $\wh{\tau}_{\mc{A}_0}$ is biased towards the treatment effect in the subpopulation for which $X = 0$.  
\vskip1ex
\noindent {\em Adjusted estimator}. In the sequel, we present a mismatch-adjusted PS estimator that accounts for the two sources of biases discussed above via two-fold re-weighting. The two
sets of weights reflect the usual propensity for treatment and additionally a propensity for 
being among the correct matches. Consider the estimator  
{\small \begin{equation}\label{eq:tau_ps-o}
\wh{\tau}_{\mc{A}}^{\text{ps-o}} = \left( \sum_{\substack{i \in \mc{A} \\ e_i = 1}} \frac{I(m_i = 0)}{1 - h_{\bm{\gamma}}(\M{z}_i)} \frac{y_i}{p_{\bm{\phi}}(\M{x}_i) } -  \sum_{\substack{i \in \mc{A} \\ e_i = 0}} \frac{I(m_i = 0)}{1 - h_{\bm{\gamma}}(\M{z}_i)} \frac{y_i}{1 - p_{\bm{\phi}}(\M{x}_i)} \right) \Big / |\mc{A}|. 
\end{equation}}The proposition below asserts unbiasedness of this estimator under the assumptions made. 
\begin{prop}\label{prop:unbiased_pso}
Under assumptions {\em {\bfseries (A1)}} through {\em {\bfseries (A5)}}, it holds that $\E[\wh{\tau}_{\mc{A}}^{ \text{{\em ps-o}}}] = \tau^*$ .  
\end{prop}
\noindent While being unbiased, $\wh{\tau}_{\mc{A}}^{ \text{{ps-o}}}$ is not a practical estimator since it requires knowledge of the parameters $\bm{\gamma}$ and $\bm{\phi}$. In the sequel, we hence outline a simple plug-in estimator using the audit sample $\mc{A}$. 
\vskip1ex
\noindent {\em Step 1}. Obtain an estimate $\wh{\bm{\gamma}}$ as follows. 
\begin{equation*}
\wh{\bm{\gamma}} = \argmin_{\bm{\gamma}} \left\{ -\sum_{i \in \mc{A}}  \{ m_i \log(h_{\bm{\gamma}}(\M{z}_i)) + (1 - m_i) \log(1 - h_{\bm{\gamma}}(\M{z}_i))  \} \right\}. 
\end{equation*}
In other words, $\wh{\bm{\gamma}}$ is the MLE under a Bernoulli model parameterized by $\bm{\gamma}$ given  the (observed) mismatch indicators in the audit sample.   
\vskip1ex
\noindent {\em Step 2}. Obtain an estimate $\wh{\bm{\phi}}$ as follows. 
\begin{equation*}
\wh{\bm{\phi}} = \argmin_{\bm{\phi}} \left\{ -\sum_{i \in \wt{\mc{A}}}  \{ e_i \log(p_{\bm{\phi}}(\M{x}_i)) + (1-e_i) \log(1 - p_{\bm{\phi}}(\M{x}_i)) \} \right\}, 
\end{equation*}
where the set $\wt{\mc{A}}$ varies depending on the three possible scenarios under consideration. Under scenario I ($\M{x}$ and $e$ contained in the same file), we may choose $\wt{\mc{A}} = \{1,\ldots,n\}$ since mismatch error will not contaminate the estimation of the PS model. In the other two scenarios, we choose $\wt{\mc{A}} = \mc{A}_0$, i.e., the set of correct matches among 
the elements of $\mc{A}$.
\vskip1ex
\noindent {\em Step 3}. Substitute $\bm{\gamma}$ and $\bm{\phi}$ in \eqref{prop:unbiased_pso} by the estimators obtained in the previous two steps.
\vskip1ex
\noindent Asymptotic standard errors of the resulting estimator $\wh{\tau}_{\mc{A}}^{\text{ps}}$ can be obtained by expressing the latter as well as $\wh{\bm{\gamma}}$ and $\wh{\bm{\phi}}$ as solutions to estimating equations. Concatenating these estimating equations, the general framework in \cite{Stefanski2002} can be applied. We refrain from spelling out specific details at this point since these will be presented when discussing more general and/or complex estimation procedures in the sequel.  

\subsection{Model-based and DR-type Estimation}\label{subsec:model-DR}
At the end of the preceding subsection, we have presented a first applicable estimator for the average treatment effect. However, this approach relies entirely on an audit sample. The audit sample is typically much smaller in size that the linked data set, which leads to poor statistical efficiency. In this subsection, we explore approaches that provide remedy in this regard. These approaches hinge on models $\mu_{\bm{\beta}}^{e}$, $e = 0,1$, for the potential outcomes, as well as on {\bfseries (A5)}, which states that for mismatched pairs of records, the two subsets of variables that are linked are assumed to be independent. For example, under Scenario I, {\bfseries (A5)}  yields $m_i = 0 \Rightarrow (\M{x}_i, e_i) \,\indep \, y_i$, $1 \leq i \leq n$. 
Regarding the outcome model, we here confine ourselves to the model
\begin{align}\label{eq:outcome_model}
\begin{split}
&Y^{(e)}|\M{x} \sim N(\mu_{\bm{\beta}}^e(\M{x}), \sigma^2), \quad \mu_{\bm{\beta}}^e(\M{x}) \coloneq \M{x}^{\T} \bm{\beta}_{\M{x}} + e \cdot \M{x}^{\T} \bm{\beta}_{e\cdot\M{x}}, \; e=0,1, \; \M{x} \in \R^p, 
\end{split}
\end{align}
with $\bm{\beta} = (\bm{\beta}_{\M{x}}^{\T}, \bm{\beta}_{e\cdot\M{x}}^{\T})^{\T}$, which amounts to using linear models with i.i.d.~Gaussian additive errors for each of the two potential outcomes; for simplicity, the intercepts are absorbed in $\M{x}$ and $\M{x} \cdot e$. These modeling assumptions simplify 
the subsequent exposition by fixing a specific example, but they are not essential to the proposed approach. In fact, it is easily seen that both the form of $\mu_{\bm{\beta}}^e$ and the error structure can be generalized within the framework for inference presented below for each of the three scenarios laid out in $\S$\ref{subsec:setup}. 

\noindent Given $\bm{\beta}$, the average treatment effect can be estimated as 
\begin{equation}\label{eq:tau_o}
\wh{\tau}^{\text{o}} = \frac{1}{n} \su \{ \mu_{\bm{\beta}}^1(\M{x}_i) - \mu_{\bm{\beta}}^0(\M{x}_i) \},
\end{equation}
which is unbiased if the model is correctly specified. The corresponding augmented propensity score (or doubly robust, DR) estimator \cite{Robins1994, Kang2007}; \cite[][$\S$13.4]{Hernan2010} is given by 
\begin{equation}\label{eq:tau_dr}
\wh{\tau}^{\text{dr}} = \wh{\tau}^{\text{o}} + \left\{ \sum_{i: \; e_i = 1} \frac{y_i - \mu_{\bm{\beta}}^1(\M{x}_i)}{p_{\bm{\phi}}(\M{x}_i)} - \sum_{i: \, e_i = 0} \frac{y_i - \mu_{\bm{\beta}}^0(\M{x}_i)}{1 - p_{\bm{\phi}}(\M{x}_i)} \right \} \Big / n,
\end{equation}
with $\wh{\tau}^{\text{o}}$ as in \eqref{eq:tau_o}. This estimator possesses the double robustness property, i.e., it is unbiased as long as not both the outcome model and the propensity score model are misspecified. Note that in the presence of mismatches, the application of these two estimators is complicated since it becomes more challenging to estimate the underlying model parameters. Moreover, 
$\wh{\tau}^{\text{dr}}$ needs to be adjusted in the same fashion as the propensity score estimators discussed in the preceding section. For what follows, we also fix the propensity score and mismatch error models as logistic regression models, i.e.,  
\begin{equation}\label{eq:prop_logistic}
p_{\bm{\phi}}(\M{x}) = \exp(\M{x}^{\T} \bm{\phi})/ \{ 1 + \exp(\M{x}^{\T} \bm{\phi})\}, \quad 
h_{\bm{\gamma}}(\M{z}) = \exp(\M{z}^{\T} \bm{\gamma})/ \{ 1 + \exp(\M{z}^{\T} \bm{\gamma})\},
\end{equation}
which are not essential assumptions, but rather serve as working examples similarly as the specific outcome model \eqref{eq:outcome_model}. As above, we assume that the intercepts are absorbed in $\M{x}$ and $\M{z}$, respectively.

\vskip2ex
\noindent {\bfseries Overarching framework}. To simplify our exposition, we suppose that the noise variance $\sigma^2$ in \eqref{eq:outcome_model} is known, noting that it is straightforward to extend our framework to estimate this quantity as well. Before discussing specifics for each of the scenarios listed in $\S$\ref{subsec:setup}, we note that these specifics follow a common pattern that can be expressed via the following set of estimation equations. 
{\small \begin{alignat}{2}\label{eq:est_eqns}
&Q_{\bm{\beta}}(\bm{\beta}, \M{m}) = \sum_{i = 1}^n Q_{i,\bm{\beta}}(\bm{\beta}, \M{m}) = \M{0}, \;\; && Q_{\bm{\gamma}}(\bm{\gamma}, \M{m}) = \sum_{i = 1}^n Q_{i,\bm{\gamma}}(\bm{\gamma},  \M{m}) = \M{0}  \\
&Q_{\bm{\phi}}(\bm{\phi}, \M{m}) = \sum_{i = 1}^n Q_{i,\bm{\phi}}(\bm{\phi}, \M{m}) = \M{0}, \;\; && Q_{\tau}(\bm{\beta}, \bm{\gamma}, \bm{\phi}, \tau, \M{m}) = \sum_{i = 1}^n Q_{i,\tau}(\bm{\beta}, \bm{\gamma}, \bm{\phi}, \tau, \M{m}) = \M{0}, \notag  \\
&Q_{\M{m}}(\bm{\beta}, \bm{\gamma}, \bm{\phi}, \M{m}) = \M{0} \; \Longleftrightarrow \, \M{m} = \big( f_i(\bm{\beta}, \bm{\gamma}, \bm{\phi}) \big)_{i = 1}^n \notag
\end{alignat}}for certain functions $\{ f_i \}$, where $\M{m} = (m_i)_{i = 1}^n$ represents the latent variables (mismatch indicators). To be clear, the above estimating equations are solved by alternating between updating all model parameters for fixed $\M{m}$ and updating the latter for fixed parameters. Let 
$\bm{\theta} = (\bm{\beta}^{\T} \, \bm{\gamma}^{\T} \bm{\phi}^{\T} \,  \tau)^{\T}$ represent
the vector of all parameters, let $Q_{\bm{\theta}}(\bm{\theta}, \M{m})$ and $Q_{i,\bm{\theta}}(\bm{\theta}, \M{m})$ be defined accordingly, and let 
\begin{align*}
J(\bm{\theta}, \M{m}) &= \begin{pmatrix}
                        \frac{\partial}{\partial \bm{\theta}} Q_{\bm{\theta}}(\bm{\theta}, \M{m}) & \frac{\partial}{\partial \M{m}} Q_{\bm{\theta}}(\bm{\theta}, \M{m}) \\[2ex]
                        \frac{\partial}{\partial \bm{\theta}} Q_{\M{m}}(\bm{\theta}, \M{m}) &  \frac{\partial}{\partial \M{m}} Q_{\M{m}}(\bm{\theta}, \M{m})
                        \end{pmatrix} \\[2ex]
                        &=  \begin{pmatrix}
                        \frac{\partial}{\partial \bm{\beta}} Q_{\bm{\beta}}(\bm{\beta}, \M{m}) & \M{0} & \M{0} &  0 & \frac{\partial}{\partial \M{m}} Q_{\bm{\beta}}(\bm{\beta}, \M{m}) \\[2ex]
                        \M{0}  &  \frac{\partial}{\partial \bm{\gamma}} Q_{\bm{\gamma}}(\bm{\gamma}, \M{m}) & \M{0} &  0 & \frac{\partial}{\partial \M{m}} Q_{\bm{\gamma}}(\bm{\gamma}, \M{m}) \\[2ex]
                         \M{0}  &  \M{0} &   \frac{\partial}{\partial \bm{\phi}} Q_{\bm{\phi}}(\bm{\phi}, \M{m}) & 0 & \frac{\partial}{\partial \M{m}} Q_{\bm{\phi}}(\bm{\phi}, \M{m}) \\[1ex]
                         \frac{\partial}{\partial \bm{\beta}} Q_{\tau}(\bm{\theta}, \M{m})  &   \frac{\partial}{\partial \bm{\gamma}} Q_{\tau}(\bm{\theta}, \M{m}) &   \frac{\partial}{\partial \bm{\phi}} Q_{\tau}(\bm{\theta}, \M{m}) & \frac{\partial}{\partial \tau} Q_{\tau}(\bm{\theta}, \M{m}) & \frac{\partial}{\partial \M{m}} Q_{\tau}(\bm{\theta}, \M{m}) \\[1ex] 
                         \frac{\partial}{\partial \bm{\beta}} Q_{\M{m}}(\bm{\theta}, \M{m})  &   \frac{\partial}{\partial \bm{\gamma}} Q_{\M{m}}(\bm{\theta}, \M{m}) &   \frac{\partial}{\partial \bm{\phi}} Q_{\M{m}}(\bm{\theta}, \M{m}) & 0 & \frac{\partial}{\partial \M{m}} Q_{\M{m}}(\bm{\theta}, \M{m})
                        \end{pmatrix}.                                                
\end{align*}
Following \cite{Elashoff2004}, the asymptotic covariance matrix of the estimator $\wh{\bm{\theta}}$ (to be introduced below) can be estimated as  
\begin{equation}\label{eq:covariance_est}
\wh{\cov}(\wh{\bm{\theta}}) =  [ \{ J(\wh{\bm{\theta}}, \wh{\M{m}}) \}^{-1} ]_{\bm{\theta} \bm{\theta}}  \; \Big \{ \sum_{i = 1}^n Q_{i,\bm{\theta}}(\wh{\bm{\theta}}, \wh{\M{m}}) \, Q_{i,\bm{\theta}}(\wh{\bm{\theta}}, \wh{\M{m}})^{\T} \Big\} \; [ \{ J(\wh{\bm{\theta}}, \wh{\M{m}}) \}^{-1} ]_{\bm{\theta} \bm{\theta}}^{\T},
\end{equation}
where the subscript $_{\bm{\theta} \bm{\theta}}$ refers to the principal submatrix corresponding to the parameters $\bm{\theta}$. While $J$ has dimension $(n+d)$-by-$(n+d)$ with $d$ denoting the dimension of $\bm{\theta}$, the desired principal submatrix can be computed efficiently (cf.~Appendix \ref{app:eff_cov}) without inverting a matrix of that dimension.  

\vskip2ex
\noindent {\bfseries Scenario I}. We have $(\M{x}, e)$ in File A and $y$ in File $B$. Accordingly, 
the propensity score model is unaffected by mismatch error, and the parameter $\bm{\phi}$ can be estimated by ordinary logistic regression, yielding the estimating equation 
\begin{align*}
Q_{\bm{\phi}}(\bm{\phi}) = \sum_{i = 1}^n \M{x}_i (e_i - p_{\bm{\phi}}(\M{x}_i)) = \M{0}
\end{align*}
with $p_{\bm{\phi}}$ as in \eqref{eq:prop_logistic}. By contrast, estimation of the outcome model 
requires an adjustment for mismatch error since the predictor variables $(\M{x}, e)$ and the outcome variable $y$ reside in separate files. We therefore adopt the likelihood-based framework in \cite{Slawski2024}, leveraging assumption {\bfseries (A5)}. Accordingly, mismatched observations are
distributed as $Y|M=1$ whose density can be shown to have  the representation (cf.~Appendix \ref{app:marginal_scenario1})
\begin{equation}\label{eq:marginal_scenario1}
f_{Y|M=1}(y) = \sum_{i = 1}^n w(\M{z}_i) \, \varphi_{\sigma}(y - \mu_{\bm{\beta}}^{e_i}(\M{x}_i)), \quad w(\M{z}_i) \coloneq \frac{h_{\bm{\gamma}}(\M{z}_i)}{\sum_{j = 1}^n h_{\bm{\gamma}}(\M{z}_j)},
%
\end{equation}
where $\varphi_{\sigma}$ denotes the PDF of the $N(0, \sigma^2)$-distribution. 
Consequently, we have the following two-component mixture model 
\begin{equation}\label{eq:mixture_scenario1}
y_i | \M{x}_i, e_i, \M{z}_i \sim (1 - h_{\bm{\gamma}}(\M{z}_i)) \varphi_{\sigma}(\cdot - \mu_{\bm{\beta}}^{e_i}(\M{x}_i)) + h_{\bm{\gamma}}(\M{z}_i) f_{Y|M=1}(\cdot), \quad 1 \leq i \leq n,
\end{equation}
which is obtained by integrating over the mismatch indicators $\{ m_i \}_{i = 1}^n$. For now, we shall assume that $f_{Y|M=1}$ is known. We can then use the Expectation-Maximization (EM) algorithm \cite{Dempster1977, Elashoff2004} to estimate the parameters $\bm{\beta}$ and $\bm{\gamma}$. The estimating equations associated with the resulting M-step are seen to be 
\begin{align*}
&Q_{\bm{\beta}}(\bm{\beta}, \wh{\M{m}}) = \sum_{i = 1}^n (1 - \wh{m}_i) \begin{bmatrix} \M{x}_i \\ e_i \cdot \M{x}_i \end{bmatrix} (y_i -\M{x}_i^{\T} \bm{\beta}_{\M{x}} - (e_i \cdot \M{x}_i)^{\T} \bm{\beta}_{e \cdot \M{x}}) = \M{0} \\
&Q_{\bm{\gamma}}(\bm{\gamma}, \wh{\M{m}}) = \sum_{i = 1}^n (\wh{m}_i - h_{\bm{\gamma}}(\M{z}_i)) = \M{0}, 
\end{align*}
where, given the  current EM iterate $(\wt{\bm{\beta}}, \wt{\bm{\gamma}})$, the $\{ \wh{m}_i \}_{i = 1}^n$ solve the estimating 
equations 
\begin{equation}\label{eq:Qm_1}
Q_{\M{m}}(\wt{\bm{\beta}}, \wt{\bm{\gamma}}, \M{m}) =  \left( \frac{f_{Y|M=1}(y_i)}{h_{\wt{\bm{\gamma}}}(\M{z}_i)  f_{Y|M=1}(y_i) + (1 - h_{\wt{\bm{\gamma}}}(\M{z}_i)) \varphi_{\sigma}(y_i - \mu_{\wt{\bm{\beta}}}^{e_i}(\M{x}_i))}  \right)_{i = 1}^n - \M{m} = \M{0},
\end{equation}
i.e., the $\{ \wh{m}_i \}_{i = 1}^n$ are given by terms inside the round brackets. Finally, the estimate for the average treatment effect
is obtained via the estimating equation
{\small \begin{align}\label{eq:Qtau_1}
Q_{\tau}(\bm{\beta}, \bm{\gamma}, \bm{\phi}, \tau, \wh{\M{m}}) &= \lambda_1 \su  (\mu_{\bm{\beta}}^{1}(\M{x}_i) - \mu_{\bm{\beta}}^{0}(\M{x}_i)) + \lambda_2 \bigg\{ \sum_{i: \, e_i = 1} (1 - \wh{m}_i) \frac{y_i - \lambda_3 \mu_{\bm{\beta}}^{1}(\M{x}_i)}{(1 - h_{\bm{\gamma}}(\M{z}_i)) p_{\bm{\phi}}(\M{x}_i)} \\
&\qquad \qquad \qquad \qquad  - \sum_{i: \, e_i = 0} (1 - \wh{m}_i) \frac{y_i - \lambda_3 \mu_{\bm{\beta}}^{0}(\M{x}_i)}{(1 - h_{\bm{\gamma}}(\M{z}_i)) (1 - p_{\bm{\phi}}(\M{x}_i))} \bigg \} \ - \, n \, \tau^{\lambda_1, \lambda_2, \lambda_3} = 0, \notag
\end{align}}where $\lambda_j$, $j=1,2,3$, can be chosen as follows: $\tau^{1,0,0}$ yields the outcome estimator \eqref{eq:tau_o}, $\tau^{0,1,0}$ yields
a (plain) propensity score estimator, and $\tau^{1,1,1}$ yields a DR-(type) estimator \eqref{eq:tau_dr}. The following statement establishes 
unbiasedness of the estimating equation \eqref{eq:Qtau_1}. 

\begin{prop}\label{prop:unbiased_est} Suppose Assumptions {\em {\bfseries (A1)}} through {\em {\bfseries (A5)}} hold true, and suppose further that the outcome
model \eqref{eq:outcome_model} and the mismatch error model $h_{\bm{\gamma}}$ are correctly specified. Then for any valid choice of 
$(\lambda_j)_{j = 1} ^3$
\begin{equation*}
\E_{\bm{\beta}, \bm{\gamma}, \bm{\phi}, \tau}[Q_{\tau}(\bm{\beta}, \bm{\gamma}, \bm{\phi}, \tau, \wh{\M{m}}(\bm{\beta}, \bm{\gamma}))] = 0,
\end{equation*}
where $\E_{\ldots}[\cdot]$ denotes the expectation when assuming that the underlying parameters $\ldots$ are given by
$\bm{\beta}, \bm{\gamma}, \bm{\phi}, \tau$, and $\wh{\M{m}}(\bm{\beta}, \bm{\gamma})$ denotes the solution defined by \eqref{eq:Qm_1}
when $\wt{\bm{\beta}} = \bm{\beta}$ and $\wt{\bm{\gamma}} = \bm{\gamma}$. 
\end{prop}
\noindent We note that Proposition \ref{prop:unbiased_est} assumes correct specification of the outcome model to ensure that the DR-type estimator satisfies an unbiased estimating equation. In this sense, that estimator does {\em not} enjoy double robustness. This aspect will be revisited in $\S$\ref{subsec:modelmis} below. 

\vskip2ex
\noindent {\bfseries Scenario II}. We have $\M{x}$ in File A and $(y, e)$ in File $B$. In this scenario, both the propensity score model and the outcome model are affected by mismatches, and we adopt a modified two-component mixture model of the following form: 
\begin{align}\label{eq:mixture_scenario2}
y_i, e_i | \M{x}_i, \M{z}_i  \sim& (1 - h_{\bm{\gamma}}(\M{z}_i)) \varphi_{\sigma}(y - \mu_{\bm{\beta}}^{e}(\M{x}_i)) \{ p_{\bm{\phi}}(\M{x}_i)^{e} \cdot (1 - p_{\bm{\phi}}(\M{x}_i))^{1-e}\} + h_{\bm{\gamma}}(\M{z}_i) f_{Y,E|M=1}(y, e), \notag\\ 
& y \in \R, \; e \in \{0,1\}, \quad 1 \leq i \leq n,
\end{align}
where the density $f_{Y,E|M=1}$ for mismatches can be expressed as (cf.~Appendix \ref{app:marginal_scenario2})
\begin{align}\label{eq:marginal_scenario2}
f_{Y,E|M=1}(y, e) &= f_{Y|E = e, M = 1}(y) \, \p(E = e|M = 1) \\
&=  \su [ \varphi_{\sigma}(y - \mu_{\bm{\beta}}^{e}(\M{x}_i))  \{ p_{\bm{\phi}}(\M{x}_i)^{e} (1 - p_{\bm{\phi}}(\M{x}_i))^{1 - e}  \} \cdot w(\M{z}_i)], \; y \in \R, e \in \{0,1\}, \notag
\end{align}
where the $\{ w(\M{z}_i) \}_{i = 1}^n$ are as in \eqref{eq:marginal_scenario1}. The estimating equations arising in the M-step when fitting the above model are given as follows.  
\begin{align}\label{eq:Qpar_2}
&Q_{\bm{\beta}}(\bm{\beta}, \wh{\M{m}}) = \sum_{i = 1}^n (1 - \wh{m}_i) \begin{bmatrix} \M{x}_i \\ e_i \cdot \M{x}_i \end{bmatrix} (y_i -\M{x}_i^{\T} \bm{\beta}_{\M{x}} - (e_i \cdot \M{x}_i)^{\T} \bm{\beta}_{e \cdot \M{x}}) = \M{0}, \\
&Q_{\bm{\phi}}(\bm{\phi}, \wh{\M{m}}) = \sum_{i = 1}^n (1 - \wh{m}_i) \M{x}_i (e_i - p_{\bm{\phi}}(\M{x}_i)) = \M{0}, \\
&Q_{\bm{\gamma}}(\bm{\gamma}, \wh{\M{m}}) = \sum_{i = 1}^n (\wh{m}_i - h_{\bm{\gamma}}(\M{z}_i)) = \M{0}, 
\end{align}
where the first and last of these estimating equations remain unchanged relative to Scenario I. Given a current EM iterate $(\wt{\bm{\beta}}, \wt{\bm{\phi}}, \wt{\bm{\gamma}})$, the $\{ \wh{m}_i \}_{i = 1}^n$ solve the estimating equations 
{\small \begin{align*}
&Q_{\M{m}}(\wt{\bm{\beta}}, \wt{\bm{\phi}}, \wt{\bm{\gamma}}, \M{m}) = \\
&\left( \frac{f_{Y,E|M=1}(y_i, e_i)}{h_{\wt{\bm{\gamma}}}(\M{z}_i) \cdot f_{Y,E|M=1}(y_i,e_i) + (1 - h_{\wt{\bm{\gamma}}}(\M{z}_i)) \varphi_{\sigma}(y_i - \mu_{\wt{\bm{\beta}}}^{e_i}(\M{x}_i)) \cdot p_{\wt{\bm{\phi}}}(\M{x}_i)^{e_i} (1 - p_{\wt{\bm{\phi}}}(\M{x}_i))^{1 - e_i}}   \right)_{i = 1}^n - \M{m} = \M{0}. 
\end{align*}}The estimating equation(s) for the ATE $\tau$ are the same as for Scenario I, cf.~\eqref{eq:Qtau_1}.
\vskip2ex
\noindent {\bfseries Scenario III}. We have $(\M{x}, y)$ in File A and $e$ in File B. As in Scenario II, mismatches can affect both the propensity score model and the outcome model. 
The two-component mixture model for this setting is of the form 
\begin{align}
y_i, e_i|\M{x}_i, \M{z}_i \sim& (1 - h_{\bm{\gamma}}(\M{z}_i)) \varphi_{\sigma}(y - \mu_{\bm{\beta}}^{e}(\M{x}_i)) \{ p_{\bm{\phi}}(\M{x}_i)^{e} \cdot (1 - p_{\bm{\phi}}(\M{x}_i))^{1-e}\} \notag \\
&+ h_{\bm{\gamma}}(\M{z}_i) \{ f_{Y|X=\M{x}_i,M=1}(y) \times \p(E = e | M = 1) \}, \; y \in \R, \, e \in \{0, 1\}, \label{eq:mixture_scenario3}
\end{align}
noting that $y_i \indep e_i | \M{x}_i$ and $e_i \indep \M{x}_i$ if $m_i = 1$, $1 \leq i \leq n$, which prompts the specific expression for the distribution of the second component. The associated terms can be shown to have the representations (cf.~Appendix \ref{app:marginal_scenario_3})
\begin{align}\label{eq:marginal_scenario3}
&f_{Y| X = \M{x}_i, M = 1}(y) = f_{Y| X = \M{x}_i}(y) = \varphi_{\sigma}(y - \mu_{\bm{\beta
}}^1(\M{x}_i)) p_{\bm{\phi}}(\M{x}_i) + \varphi_{\sigma}(y - \mu_{\bm{\beta
}}^0(\M{x}_i)) (1 - p_{\bm{\phi}}(\M{x}_i)), \; y \in \R,  \notag \\
&\p(E = e| M = 1) = \su \{ p_{\bm{\phi}}(\M{x}_i)^e \cdot (1 - p_{\bm{\phi}}(\M{x}_i))^{1-e} \} w(\M{z}_i), \; e \in \{0,1\}, 
\end{align}
where the $\{ w(\M{z}_i) \}_{i = 1}^n$ are as in \eqref{eq:marginal_scenario1}. The estimating equations arising in the M-step when fitting the above model have the same form as \eqref{eq:Qpar_2} for $\bm{\beta}$, 
$\bm{\phi}$ and $\bm{\gamma}$. Using \eqref{eq:marginal_scenario3}, the estimating equation for $\M{m}$ becomes 
{\small \begin{align*}
&Q_{\M{m}}(\wt{\bm{\beta}}, \wt{\bm{\phi}}, \wt{\bm{\gamma}}, \M{m}) = \\
&\hspace*{-1.5ex}\bigg( \frac{f_{Y|X=\M{x}_i}(y_i) \cdot \p(E = e_i | M = 1)}{h_{\wt{\bm{\gamma}}}(\M{z}_i) \cdot f_{Y|X=\M{x}_i}(y_i) \cdot \p(E = e_i | M = 1) + (1 - h_{\wt{\bm{\gamma}}}(\M{z}_i)) \cdot \varphi_{\sigma}(y_i - \mu_{\wt{\bm{\beta}}}^{e_i}(\M{x}_i)) \cdot p_{\wt{\bm{\phi}}}(\M{x}_i)^{e_i} (1 - p_{\wt{\bm{\phi}}}(\M{x}_i))^{1 - e_i}}   \bigg)_{i = 1}^n \hspace*{-1ex} \\
&- \M{m} = \M{0}, 
\end{align*}} where, as previously, $(\wt{\bm{\beta}}, \wt{\bm{\phi}}, \wt{\bm{\gamma}})$ represent a current EM iterate for the parameters. 

\subsection{Model Misspecification}\label{subsec:modelmis}
In this subsection, we discuss aspects related to model misspecification. The approach outlined 
in the previous subsection involves the following three models. 
\begin{itemize}
\item[(M1)] The model for the mismatch indicators $h_{\bm{\gamma}}$, 
\item[(M2)] The outcome models $\mu_{\bm{\beta}}^e$, $e=0,1$,  
\item[(M3)] The propensity score model $p_{\bm{\phi}}$. 
\end{itemize}
In addition, the mixture model components for mismatched units (cf.~\eqref{eq:marginal_scenario1}, \eqref{eq:marginal_scenario2}, \eqref{eq:marginal_scenario3}) might be subject to misspecification as well. The impact of such misspecification is comparable to that of a violation of the outcome model (M2).     

\vskip1ex
\noindent {\em Impacts of model misspecification}. Among the three, violation of (M3) is the least critical. In Scenario I estimation of the average treatment effect via $\wh{\tau}^{\lambda_1, \lambda_2, \lambda_2}$ based on the estimating equation \eqref{eq:Qtau_1} yields consistency of $\wh{\tau}^{1,0,0}$ (plain outcome estimator) and $\wh{\tau}^{1,1,1}$ (DR-type estimator); only the plain propensity score estimator $\wh{\tau}^{0,1,0}$ is generally inconsistent. In Scenario II, consistency can be maintained by reducing the mixture model \eqref{eq:mixture_scenario2} as follows: 
\begin{equation*}
y_i| \M{x}_i, \M{z}_i  \sim (1 - h_{\bm{\gamma}}(\M{z}_i)) \varphi_{\sigma}(y - \mu_{\bm{\beta}}^{e}(\M{x}_i)) + h_{\bm{\gamma}}(\M{z}_i) f_{Y|M=1}(y), \;\;  y \in \R, \quad 1 \leq i \leq n,
\end{equation*}
where accordingly $f_{Y|M=1}$ is given by 
\begin{equation*}
f_{Y|M=1}(y) = \su [ \varphi_{\sigma}(y - \mu_{\bm{\beta}}^{e}(\M{x}_i)) \cdot w(\M{z}_i)], \quad y \in \R. 
\end{equation*}
The situation in Scenario III is not quite as straightforward since the $\{ f_{Y|X = \M{x}_i}(\cdot) \}_{i = 1}^n$ depend on the underlying PS model, but this may not pose (much of) an issue 
if these quantities can be specified (approximately) correctly without direct reference to that model. Apart from that, one proceeds as for Scenario II, i.e., the mixture model 
is reduced to a model for $y_i | \M{x}_i, \M{z}_i$, $1 \leq i \leq n$. 

Violation of (M2) is much more critical since the outcome model plays a crucial role in the estimation of the $\{ \wh{m}_i \}_{i = 1}^n$, which contribute to all three estimators under consideration. Consequently, all three estimators are generally inconsistent, i.e., none of them enjoys robustness with respect to model misspecification. The consequence of a violation of (M3)
is even more severe since additionally, the PS approach based on an audit sample \eqref{eq:tau_ps-o} will incur bias, regardless of whether (M1) holds true.
\vskip1ex
\noindent {\em Potential improvements of PS estimators based on an audit sample}. In the sequel, we shall assume that (M1) and (M3) hold true, while (M2) is (moderately) misspecified. In this situation, we propose the use of the DR-(type) estimator restricted to the audit sample, using the full sample to estimate the parameters of the (misspecified) outcome model. Since (M3) is assumed to be true, the resulting estimator of the ATE is consistent and expected to achieve substantially smaller variance in comparison to the plain PS estimator $\wh{\tau}_{\mc{A}}^{\text{ps}}$ \cite{Lunceford2004}. We recall that the latter is derived from \eqref{eq:tau_ps-o} by substituting the parameters $\bm{\gamma}$ and $\bm{\phi}$ with their estimates. The augmented estimator $\wh{\tau}_{\mc{A}}^{\text{dr}}$ is given by\\
\begin{minipage}{\textwidth} 
{\small \begin{equation}\label{eq:est_dr_audit}
\wh{\tau}_{\mc{A}}^{\text{dr}} = \frac{1}{n} \sum_{i=1}^n \{ \mu_{\wh{\bm{\beta}}}^1(\M{x}_i) - \mu_{\wh{\bm{\beta}}}^0(\M{x}_i) \} +  \Bigg( \sum_{\substack{i \in \mc{A} \\ e_i = 1}} \frac{I(m_i = 0)}{1 - h_{\wh{\bm{\gamma}}}(\M{z}_i)} \frac{y_i - \mu_{\wh{\bm{\beta}}}^{1}(\M{x}_i)}{p_{\wh{\bm{\phi}}}(\M{x}_i) } -  \sum_{\substack{i \in \mc{A} \\ e_i = 0}} \frac{I(m_i = 0)}{1 - h_{\wh{\bm{\gamma}}}(\M{z}_i)} \frac{y_i - \mu_{\wh{\bm{\beta}}}^{0}(\M{x}_i)}{(1 - p_{\wh{\bm{\phi}}}(\M{x}_i)} \Bigg) \Big / |\mc{A}|,    
\end{equation}}
\end{minipage}\\[1ex]
where the parameter estimates $\wh{\bm{\beta}}$, $\wh{\bm{\gamma}}$ and $\wh{\bm{\phi}}$ are obtained 
as the solutions of the following estimating equations. 
\begin{align}
&Q_{\bm{\gamma}}(\bm{\gamma}) = \sum_{i \in \mc{A}} \M{z}_i (m_i - h_{\bm{\gamma}}(\M{z}_i)) = 0, \label{eq:gammahat_misspec}\\
&Q_{\bm{\phi}}(\bm{\phi}, \wh{\M{m}}_{\bm{\phi}}) =\sum_{i = 1}^n (1 - \wh{m}_{i,\bm{\phi}})  \M{x}_i (e_i - p_{\bm{\phi}}(\M{x}_i)), \notag \\
&Q_{\bm{\beta}}(\bm{\beta}, \wh{\M{m}}_{\beta}) = \sum_{i = 1}^n (1 - \wh{m}_{i,\bm{\beta}}) \begin{bmatrix} \M{x}_i \\ e_i \cdot \M{x}_i \end{bmatrix} (y_i -\M{x}_i^{\T} \bm{\beta}_{\M{x}} - (e_i \cdot \M{x}_i)^{\T} \bm{\beta}_{e \cdot \M{x}}), \notag 
\end{align}
where we maintain two sets of estimates of the mismatch indicators, 
$\{ \wh{m}_{i,\bm{\phi}}\}$ and $\{ \wh{m}_{i,\bm{\beta}}\}$, which are obtained 
according to the E-steps associated with the latter two estimating equations (note that $\wt{\bm{\phi}}$, $\wt{\bm{\beta}}$  are updated iteratively, while $\wh{\bm{\gamma}}$ is obtained directly from \eqref{eq:gammahat_misspec}):\\ 
\begin{minipage}{\textwidth}
{\small \begin{align*}
&Q_{\M{m}_{\bm{\phi}}}(\wt{\bm{\phi}}, \wh{\bm{\gamma}}, \M{m}_{\bm{\phi}}) =   \left( \frac{f_{E|M=1}(e_i)}{h_{\wh{\bm{\gamma}}}(\M{z}_i)  f_{E|M=1}(e_i) + (1 - h_{\wh{\bm{\gamma}}}(\M{z}_i)) \cdot p_{\wt{\bm{\phi}}}(\M{x}_i)^{e_i} \cdot (1 - p_{\wt{\bm{\phi}}}(\M{x}_i))^{1 - e_i}}   
\right)_{i = 1}^n - \M{m}_{\bm{\phi}} = \M{0}, \\
&Q_{\M{m}_{\bm{\beta}}}(\wt{\bm{\beta}}, \wh{\bm{\gamma}}, \M{m}_{\bm{\beta}}) =   \left( \frac{g(y_i, e_i, \M{x}_i)}{h_{\wh{\bm{\gamma}}}(\M{z}_i)  g(y_i, e_i, \M{x}_i) + (1 - h_{\wh{\bm{\gamma}}}(\M{z}_i)) \cdot   
\varphi_{\sigma}(y_i - \mu_{\wt{\bm{\beta
}}}^{e_i}(\M{x}_i))} \right)_{i = 1}^n - \M{m}_{\bm{\beta}} = \M{0}. 
\end{align*}}
\end{minipage}\\[2ex]
Furthermore, we set $\wh{m}_{i, \bm{\phi}} = \wh{m}_{i, \bm{\beta}} = m_i$ for all 
$i \in \mc{A}$; in Scenario I, $\wh{m}_{i,\bm{\phi}} \equiv 0$. In the above display,
$g(y_i, e_i, \M{x}_i)$ is a placeholder for the following quantities.  
\begin{align}\label{eq:marginal_scenario2_audit}
&\text{\underline{Scenario I}:}\notag\\
&f_{Y|M=1}(y_i) = \sum_{j = 1}^n w(\M{z}_j) \, \varphi_{\sigma}(y_i - \mu_*(\M{x}_j, e_j)), \quad w(\M{z}_j) = \frac{h_{\bm{\gamma}}(\M{z}_j)}{\sum_{k = 1}^n h_{\bm{\gamma}}(\M{z}_k)}, \;\;\, 1 \leq j \leq n, \notag \\[1ex]
&\text{\underline{Scenario II}:}\notag\\
&f_{Y|E=e_i, M=1}(y_i) = \sum_{j = 1}^n w_{e_i}(\M{z}_j, \M{x}_j) \varphi_{\sigma}(y_i - \mu_*(\M{x}_j, e_i)), \\
&\qquad \quad w_e(\M{z}_j, \M{x_j}) = \frac{h_{\bm{\gamma}}(\M{z}_j) \cdot \{ p_{\bm{\phi}}(\M{x}_j) \}^{e} \{ 1 - p_{\bm{\phi}}(\M{x}_j) \}^{1 - e}}{\sum_{k = 1}^n [h_{\bm{\gamma}}(\M{z}_k) \cdot \{ p_{\bm{\phi}}(\M{x}_k) \}^{e} \{ 1 - p_{\bm{\phi}}(\M{x}_k) \}^{1 - e}] }, \; \; e \in \{0,1\}, \;\;\, 1 \leq j \leq n, \notag\\[1ex]  
&\text{\underline{Scenario III}:} \notag\\
&f_{Y|X=\M{x}_i, M=1}(y_i) = \varphi_{\sigma}(y - \mu_*(\M{x}_i, 1)) p_{\bm{\phi}}(\M{x}_i) + \varphi_{\sigma}(y - \mu_{*}(\M{x}_i, 0)) (1 - p_{\bm{\phi}}(\M{x}_i)), \notag
\end{align}
for $i=1,\ldots,n$, where $\mu_*(\M{x}, e)$ represents the true outcome model; for simplicity, we suppose that the Gaussian model in \eqref{eq:outcome_model} continues to hold apart from the change regarding $\mu$.  The expressions for Scenarios I and III are analogous to those in \eqref{eq:marginal_scenario1} and \eqref{eq:marginal_scenario3}. The expression for Scenario II is derived in Appendix \ref{app:marginal_extra}. 

\section{Simulations}\label{sec:simulation}
The following section presents the results of simulations conducted to evaluate the empirical performance of the approach. These can be roughly divided into two parts. In the first part, we assume that all models are correctly specified. In the second part, we consider two different forms of misspecifications associated with the mixture
models used for the outcome, assuming the presence of an audit sample. All three 
scenarios described in $\S$\ref{subsec:setup}  are examined in each case. 

\subsection{Correct Model Specification}
{\em Setup}. We consider $X \sim U(0, 3)$, and set $Z = X$, i.e., the same covariate
is used for the outcome/propensity score and mismatch error model, respectively. The latter are specified as follows: 
\begin{alignat}{2}
&E|X=x \sim \mathrm{Bernoulli}(\{1 + \exp(-\phi_0^* - \phi^* x) \}^{-1}), \quad &&\phi_0^* = -2, \; \phi^* = 1, \notag \\
&M|X=x \sim \mathrm{Bernoulli}(\{1 + \exp(-\gamma_0^* - \gamma^* x) \}^{-1}), \quad &&\gamma_0^* = -10, \; \gamma_1^* = 5, \label{eq:simualtion_model}\\
&Y|E=e, X=x \sim N(\beta_0^* + \beta_e^* \cdot e + \beta_x^* \cdot x + \beta_{e \cdot x}^* \cdot x, 1), \quad &&\beta_0^* = 3, \;  \beta_e^* = 1.5, \; \beta_x^* = 2, \beta_{e \cdot x}^* = 1. \notag 
\end{alignat}
It is easy to verify that under the above outcome model and the distribution of $X$, we have $\tau^* = 3$ for the average treatment, which is the quantity of interest. Furthermore, the expected fraction of treated observations equals about $.395$ and the expected rate of mismatched observations equals about $1/3$.

Data is generated as follows: we first sample $X$ and then generate the remaining variables according to the setup above, yielding $\{ (x_i, e_i, m_i, y_i) \}_{i = 1}^n$ with $n=1,000$; we consider reasonably large sample sizes throughout since linked datasets, particularly in applications such as electronic health records, tend to be even larger. In Scenario I, mismatches are introduced by applying a permutation of maximum cycle length to the subsets of 
$y$'s for which the corresponding $m$ is equal to one. In Scenario III, this process is applied to the $e$'s instead of the $y$'s, and in Scenario II, $(y,e)$-pairs are permuted jointly.
\vskip2ex
\noindent {\em Results}. In Table \ref{tab:sim:correctmodel}, we report bias, standard deviation, and coverage of confidence intervals of several estimators of the ATE, including the naive propensity score estimator $\wh{\tau}_{\text{naive}}$ (assuming knowledge of the underlying PS model) and the estimators $\wh{\tau}^{\text{o}} = \wh{\tau}^{1,0,0}$ (outcome estimator), 
$\wh{\tau}^{\text{ps}} = \wh{\tau}^{0,1,0}$ (PS estimator based on the full data set), $\wh{\tau}^{\text{dr}} = \wh{\tau}^{1,1,1}$ (DR-type estimator) as defined in \eqref{eq:Qtau_1}. Estimation 
of the parameters $\bm{\beta}^*= (\beta_0^*, \beta_e^*, \beta_x^*, \beta_{e \cdot x}^*)$, $\bm{\gamma}^* = (\gamma_0^*, \gamma_1^*)$, and $\bm{\phi}^* = (\phi_0^*, \phi^*)$ is based on fitting the scenario-specific mixture models as described in
the accordingly labeled paragraphs in $\S$\ref{subsec:model-DR} using the EM algorithm. For computational simplicity, the variance in the outcome model and the mixture model components associated with the mismatches are assumed known. We also consider the ``oracle", which is the outcome estimator using a dataset containing all observations in their correct pairing (i.e., there are no mismatches). 

The numbers in Table \ref{tab:sim:correctmodel} confirm that the naive propensity 
score estimator can exhibit substantial bias due to the selection induced
by restricting the analysis to correctly matched pairs only. The bias of the outcome model-based $\wh{\tau}_{\text{o-ig}}$ and PS $\wh{\tau}_{\text{ps-ig}}$ estimators ignoring mismatches is even larger. The (adjusted) outcome and DR-type estimators maintain appropriate coverages in all three scenarios and achieve comparable efficiencies, with $\wh{\tau}^{\text{o}}$ having a slight edge over $\wh{\tau}^{\text{dr}}$. Compared to the oracle estimator that is equipped with the correctly linked data, standard errors are about a factor of 1.6 higher. The adjusted propensity score estimator $\wh{\tau}^{\text{ps}}$ has consistently low bias, but significantly larger standard errors, which is expected. It exhibits over-coverage in Scenarios I and II, and slight under-coverage in Scenario III. Overall, the results are promising in that they show that the usual estimators of the ATE can be adjusted to counter bias that arises from the presence of mismatched records.   
\begin{table}
\centering

{\small \begin{tabular}{lccc|ccc|ccc}
\toprule
\multirow{2}*{} & \multicolumn{3}{c|}{{\bfseries Scenario I}} & \multicolumn{3}{c|}{{\bfseries Scenario II}} & \multicolumn{3}{c}{{\bfseries Scenario III}} \\
                        & $|\text{Bias}|$ & SD & CVG & $|\text{Bias}|$  & SD & CVG & $|\text{Bias}|$  & SD & CVG \\
\midrule
$\wh{\tau}_{\text{o-ig}}$  & 1.18 & 0.126 & --/-- & 0.28 & 0.098 & --/--  & 1.28 & 0.112 & --/-- \\
$\wh{\tau}_{\text{ps-ig}}$  & 1.31 & 0.139 & --/-- & 2.61 & 0.164 & --/--  & 2.61 & 0.164 & --/-- \\
$\wh{\tau}_{\text{naive}}$ & 0.48 & 0.601 & --/-- & 0.48 & 0.602 & --/--  & 0.48 & 0.602 & --/-- \\
$\wh{\tau}^{\text{ps}}$ & 0.03 & 0.352 & 99.5\% & 0.01 & 0.407 & 98.9\% & 0.00 & 0.545 & 90.6\% \\
$\wh{\tau}^{\text{o}}$ & 0.00 & 0.122 & 95.3\% & 0.01 & 0.127 & 96.0\% & 0.00 & 0.124 & 95.0\% \\
$\wh{\tau}^{\text{dr}}$  & 0.01 & 0.126 & 96.3\% & 0.01 & 0.133 & 95.4\% & 0.00 & 0.131 & 95.2\% \\
{\footnotesize oracle}  & 0.00 & 0.078 & 94.7\% & 0.00 & 0.078 & 94.4\% & 0.00 & 0.078 & 94.5\% \\
\bottomrule
\end{tabular}\\[.5ex]
$\wh{\tau}_{\text{o-ig}}$, $\wh{\tau}_{\text{ps-ig}}$: conventional outcome model-based  and PS estimators ignoring mismatches.\\
--/--: Not evaluated because appropriate coverage levels are not expected. 
}
\vspace*{-.5ex}
\caption{Absolute bias, standard deviation (SD), and confidence interval coverage (CVG) for the estimators of the average treatment effect described in the text in the setting of correct model specification. The numbers represent averages over 1,000 independent replications.}\label{tab:sim:correctmodel}
\end{table}

\subsection{Model misspecification}
In the first set of simulations involving model misspecification, we maintain the models in \eqref{eq:simualtion_model} with the exception of the outcome model, which is changed as follows 
\begin{align*}
&Y|E=0, X = x \sim N\left(\beta_0^* + \beta_x^* - \frac{1}{4} (x^2 +  |\sin(2\pi\cdot x/3)|, 1 \right)  \\
&Y|E=1, X = x \sim N\left(\beta_0^* + \beta_e^* +  (\beta_x^* + \beta_{x \cdot e}^*)  (\exp(0.3 \cdot (x - \sqrt{x}))), 1 \right), 
\end{align*}
where the coefficients $\beta_0^*, \beta_e^*, \beta_x^*, \beta_{x \cdot e}^*$ are as in \eqref{eq:simualtion_model}. Under the above model, we have $\tau^* \approx 2.452$ for the ATE. 

In a second set of simulations, we adopt the outcome model in 
\eqref{eq:simualtion_model}, but misspecify the second component 
(corresponding to mismatches) in the scenario-specific mixture models 
\eqref{eq:mixture_scenario1}, \eqref{eq:mixture_scenario2}, \eqref{eq:mixture_scenario3}. In Scenario I, \eqref{eq:marginal_scenario1} is misspecified as $f_Y$ (i.e., the marginal density of the entire $Y$'s) instead of $f_{Y|M=1}$. In Scenario II, $f_{Y,E|M=1}$ \eqref{eq:marginal_scenario2} is misspecified as $f_{Y,E}$. In Scenario III, we fit 
separate mixture models for $e_i | \M{x}_i, \M{z}_i$ and 
$y_i | \M{x}_i, \M{z}_i$, $1 \leq i \leq n$; from the former, 
we obtain an estimate $\wh{\bm{\phi}}$, which is then substituted
for $\bm{\phi}$ when evaluating $f_{Y|X=\M{x}_i,M}$ in \eqref{eq:marginal_scenario1}. 

For both sets of simulations, we generate $n=10,000$ samples in total out of which $1,000$ are assigned to an audit sample $\mc{A}$ for which 
the associated mismatch indicators $\{ m_i \}$ are considered known. We then adopt the approach described in the second paragraph of $\S$\ref{subsec:modelmis}. In particular, we compare
$\wh{\tau}_{\mc{A}}^{\text{ps}}$, i.e., the audit sample-based propensity score estimator \eqref{eq:tau_ps-o} with $\bm{\phi}$ and $\bm{\gamma}$ substituted by estimates, as well as the audit-sample based DR-type estimator $\wh{\tau}_{\mc{A}}^{\text{dr}}$ \eqref{eq:est_dr_audit}. We also study the outcome-based estimator $\wh{\tau}^{\text{o}}$, the PS estimator $\wh{\tau}^{\text{ps}} = \wh{\tau}^{0,1,0}$ and the DR-type estimator $\wh{\tau}^{\text{dr}}$ based on the full data set; in the presence of model misspecification, these estimators are generally subject to bias.
\vskip2ex
\noindent {\em Results}. The top part of Table \ref{tab:sim:misspec} displays the results 
of the set of simulations in which the outcome model is misspecified. As expected, the estimators $\wh{\tau}^{\text{ps}}$, $\wh{\tau}^{\text{o}}$, and $\wh{\tau}^{\text{dr}}$, which rely on the imputation of the mismatch indicators $\{ m_i \}$, exhibit a notable bias that ranges between 6\% and 14\%. The two estimators based on the audit-sample are essentially unbiased and achieve close to nominal coverage, but exhibit significantly larger variation. Note that the DR-type estimator $\wh{\tau}_{\mc{A}}^{\text{dr}}$ achieves more than a sixfold reduction in standard deviation compared to $\wh{\tau}_{\mc{A}}^{\text{ps}}$. The second set of simulations in the bottom part of Table \ref{tab:sim:misspec} show that $\wh{\tau}^{\text{o}}$ and $\wh{\tau}^{\text{dr}}$
are largely robust to the second type of model misspecification in the sense that the bias and under-coverage tend to be moderate. We observe that in Scenario II, the misspecification under consideration affects the estimation of the PS model, leading to substantial bias of $\wh{\tau}_{\mc{A}}^{\text{ps}}$, whereas the bias of $\wh{\tau}_{\mc{A}}^{\text{dr}}$ is negligible.   
\begin{table}[ht]
\centering
\textsf{Set 1: Misspecification of the outcome model}\\[.5ex]
{\small \begin{tabular}{lccc|ccc|ccc}
\toprule
\multirow{2}*{} & \multicolumn{3}{c|}{{\bfseries Scenario I}} & \multicolumn{3}{c|}{{\bfseries Scenario II}} & \multicolumn{3}{c}{{\bfseries Scenario III}} \\
                        & $|\text{Bias}|$ & SD & CVG & $|\text{Bias}|$  & SD & CVG & $|\text{Bias}|$  & SD & CVG \\
\midrule
$\wh{\tau}^{\text{ps}}$ & 0.21 & 0.174 & --/-- & 0.22 & 0.287 & --/-- & 0.35 & 0.289 & --/-- \\
$\wh{\tau}^{\text{o}}$ & 0.24 & 0.040 & --/--  & 0.20 &  0.040 & --/-- & 0.25 & 0.041 & --/-- \\
$\wh{\tau}^{\text{dr}}$  & 0.19 & 0.050 & --/--  & 0.15 & 0.052 & --/-- & 0.20 & 0.051 & --/-- \\
$\wh{\tau}_{\mc{A}}^{\text{ps}}$  & 0.04 & 1.716 & 93.6\% & 0.04 & 1.695 & 93.6\% & 0.04 & 1.695 & 93.6\% \\
$\wh{\tau}_{\mc{A}}^{\text{dr}}$  & 0.03 & 0.285 & 92.4\% & 0.03 & 0.281 & 93.0\% & 0.03 & 0.283 & 92.9\% \\
\bottomrule
\end{tabular}}
\vskip5ex
\textsf{Set 2: Misspecification of the second mixture component}\\[.5ex]
{\small \begin{tabular}{lccc|ccc|ccc}
\toprule
\multirow{2}*{} & \multicolumn{3}{c|}{{\bfseries Scenario I}} & \multicolumn{3}{c|}{{\bfseries Scenario II}} & \multicolumn{3}{c}{{\bfseries Scenario III}} \\
                        & $|\text{Bias}|$ & SD & CVG & $|\text{Bias}|$  & SD & CVG & $|\text{Bias}|$  & SD & CVG \\
\midrule
$\wh{\tau}^{\text{ps}}$ & 3.49 & 0.396 & --/-- & 1.78 & 0.438 & --/-- & 0.21 & 0.558 & --/-- \\
$\wh{\tau}^{\text{o}}$ & 0.01 & 0.037 & 89.7\%$^*$  & 0.04 &  0.040 & 81.0\%$^*$ & 0.01 & 0.041 & 86.0\%$^*$ \\
$\wh{\tau}^{\text{dr}}$  & 0.01  & 0.048 & 97.8\%$^*$  & 0.02 & 0.055 & 95.3\%$^*$ & 0.00 & 0.079 & 64.4\%$^*$ \\
$\wh{\tau}_{\mc{A}}^{\text{ps}}$  & 0.00 & 2.299  & 93.9\% & 0.80 & 2.420 & 93.2\% & 0.00 & 2.270 & 94.1\% \\
$\wh{\tau}_{\mc{A}}^{\text{dr}}$  & 0.01 & 0.231 & 97.1\% & 0.03 & 0.281 & 98.5\% & 0.01 & 0.225 & 97.9\% \\
\bottomrule
\end{tabular}
\\[.5ex]
--/--: Not evaluated because appropriate coverage levels are not expected.\\
$^{\star}$: Evaluated, but appropriate coverage may not be attained given the impact of misspecification.
}
\caption{Absolute bias, standard deviation (SD), and confidence interval coverage (CVG) for the estimators of the average treatment effect described in the text in the presence of two types of model misspecification.}\label{tab:sim:misspec}
\end{table}

\section{Case Study}\label{sec:casestudy}
\begin{figure}
\begin{center}
\includegraphics[height = 0.17\textheight]{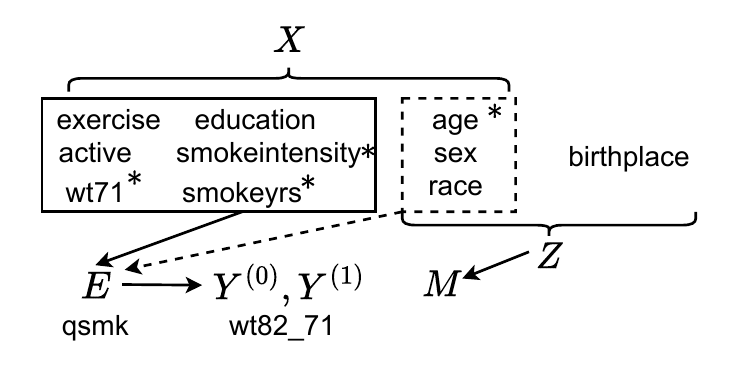} 
\end{center}
\vspace*{-2ex}
\caption{Overview on the variables (and their roles) used in the analysis of the case study. Asterisked variables enter the PS and outcome model in terms of a quadratic function.}\label{fig:nhefs_diagram}
\end{figure}

We here analyze data from the NHEFS (National Health and Nutrition Examination Survey), modifying
the analysis in the causal inference textbook by Hernan \& Robins \cite{Hernan2010}.  The data 
set and R code of the original analysis is available from the companion webpage \cite{Hernan2025}.
The goal of the analysis is to investigate the effect of smoking cessation \textsf{qsmk} ($E$, $1$: yes, $0$: no) on weight gain \textsf{wt82\_71} ($Y$, in kg) -- the numbers 82 and 71 refer to the years at baseline (1971) and follow-up (1982), respectively. Potential confounder variables $X$ include \textsf{age}, \textsf{sex}, and \textsf{race} (white yes/no), number of cigarettes 
consumed per day (\textsf{smokeintensity}), years of smoking prior to possible cessation (\textsf{smokeyrs}), the weight at the baseline (\textsf{wt71}), \textsf{education} (a five-level
ordered factor with $5$ corresponding to a the highest level, a college degree),
\textsf{exercise} (a three-level ordered factor quantifying the extent of physical exercise), 
and \textsf{active} (a three-level ordererd factor quantifying everyday activity). Study participants
corresponding to $E = 1$ and $E = 0$, respectively, tend to exhibit some notable difference
with regard to these characteristics (cf.~\cite[][Table 12.1]{Hernan2010}). A summary of variables
under consideration is provided in the diagram in Figure \ref{fig:nhefs_diagram}.  

The following outcome and PS models are used in \cite{Hernan2010}:  
\begin{align*}
&Y|E=e, \, X=x \sim N(\eta(x) + e, \sigma^2),  \quad \; E=e| X=x \sim \mathrm{Bernoulli} \left( \frac{\exp(\eta(x))}{1 + \exp(\eta(x))} \right)\\
&\eta(x) \coloneq  \text{Intercept} + q(\textsf{age}) + \textsf{sex} + \textsf{race} + q(\textsf{smokeyrs}) + q(\textsf{smokeintensity}) + \text{cat}(\textsf{education}) \\ 
&\qquad + q(\textsf{wt\_71}) + \text{cat}(\textsf{active}) + \text{cat}(\textsf{exercise})
\end{align*}
where $q(\cdot)$ refers to a quadratic function in the respective variables and $\text{cat}(\cdot)$ emphasizes that the variable enters the model as a categorical (i.e., factor) variable and is coded accordingly; furthermore, it is understood that the (suppressed) parameters associated with each term are different across the outcome and PS model. 

We investigate the impact of linkage errors on the above analysis under Scenario II, 
i.e., the outcome $Y$ and treatment status $E$ are contained in the same file while 
$X$ is linked from a second file, we simulate match status $M$ according to the logistic
regression model $\p(M = 1 | X = x) = \exp(\eta_{\bm{
\gamma}}(x)) / \{ 1 + \exp(\eta_{\bm{
\gamma}}(x)) \}$, where $\eta_{\bm{\gamma}}(x) = \gamma_0 + \gamma_1 \cdot \textsf{age} + 
\gamma_2 \cdot \textsf{sex} + \gamma_3 \textsf{race} + \gamma_4 \cdot \textsf{bp\_freq\_col}$,
where $\bm{\gamma} = (\gamma_0, \ldots, \gamma_4)^{\T} = (2, -0.1, .75, 1.2, .5)^{\T}$ and then
randomly permute records for which $M = 1$; with this model, the overall mismatch
rate is around 15\%. This model can be loosely motivated as follows. Record linkage is often based on quasi-identifiers such as names, date of birth, and residential address. Older subjects tend to have a more steady lifestyle, which reduces the chance of changes in name or address. Female 
subjects (\textsf{sex}=1) adopt their spouse's name in about 80\% of cases, which reduces the reliability of surname as quasi-identifier. For non-white subjects (\textsf{race}=1), in particular for many minority populations such as Hispanic individuals, linkage is often 
more challenging \cite{abowd2021}, e.g., because name recording and comparisons during linkage tend to be geared towards English names. Finally, \textsf{bp\_col\_freq} is defined
as the log of the birthplace (i.e.,~US states) frequency (divided by the maximum 
frequency over all states). This variable is used as a surrogate for the frequency of some form of place of residence/address, and is -- unlike all other variables in the model for $M$ -- considered unrelated to $E$ and $Y$. 
\vskip2ex
\noindent {\bfseries Results}. Table \ref{tab:casestudy} displays various estimates 
for the ATE, with and without mismatch error. In the latter case, the ``plain" ATE that    
disregards potential confoundedness yields an ATE of 2.54 (weight gain in kg). PS, outcome, and the DR estimator yield notably different ATEs that are close to 
each other, with a range from 3.42 to 3.46, which is considered the benchmark when evaluating
various estimators in the presence of mismatch error. Since the latter is introduced 
randomly, we report averages over 100k independent replications. First, we consider
PS and outcome model-based estimators that confine the analysis to the set of 
correctly linked observations. These are not fully practical estimators since
the match status for each observation is generally unknown, but they mimic the
commonly adopted strategy of only using those observations that are deemed
``safe matches", i.e., for which the probability of a mismatch is considered negligible. For the reasons described in $\S$\ref{subsec:PSest}, these estimators are labeled ``naive" since they
do not account for the potential selection bias introduced in this way. This bias also manifests
in the case study under consideration even though it is not dramatic, with estimates
of 3.55 (PS estimator) and 3.52 (outcome model estimator) slightly larger than the $[3.42, 3.46]$ range. On the other hand, we consider estimators ignoring linkage error altogether, i.e., they assume that each record in the linked data set is a correct link. The PS estimate
obtained in this way is rather close to the original result ($3.41$ vs.~$3.42$), which does 
not come as a surprise: note that since outcomes $y$ and treatment status $e$ are contained 
in the same file by construction, mismatches affect the alignment of propensity scores and exposure status as well as the estimation of the PS model, which 
is generally less impacted since (i) not all mismatches lead to a change in treatment status, (ii) the contamination introduced by mismatches is limited since it leads to swaps of zeroes and ones only for a fraction of the observations that is less than the overall 
mismatch rate of about 15\%. By contrast, the impact on the outcome model-based estimator is 
more severe, leading to a marked drop in the ATE estimate ($3.32$) relative to the range
$[3.42, 3.46]$ used as the benchmark. Next, we consider estimators that account for mismatches according to the approaches laid out in $\S$\ref{subsec:model-DR}, equipped with the correct model for the match status $M$. We note that the adjusted
outcome model-based estimator is only partially successful in restoring the estimate obtained,
on the mismatch-free data, with a change from $3.32$ (no adjustment) to $3.36$. By contrast, the 
adjusted PS and DR-type estimators equal $3.45$, which is well within the benchmark range. The varied degree of success can be explained as follows: the outcome-model based estimator relies
directly on the restoration of the original regression parameter estimate, whereas the two other
estimators require a proper weighing of the observations in a manner that mismatches 
are down-weighed and correctly matched observations are reweighed to address potential selection bias. Finally, we note that the estimators based on a randomly selected audit sample of size 10\% are not suitable because their variability is dramatically larger than those of the other estimators. 
\begin{table}
\addtolength{\tabcolsep}{-.5pt}   
{\small \begin{tabular}{|l|llll|lllllllll|}
\hline &&&&&&&&&&&&&\\[-1ex]
         & $\wh{\tau}_*^{\text{pl}}$ & $\wh{\tau}_*^{\text{ps}}$ &  $\wh{\tau}_*^{\text{o}}$ &$\wh{\tau}_*^{\text{dr}}$ & $\wh{\tau}_{\text{naive}}^{\text{ps}}$ & $\wh{\tau}_{\text{naive}}^{\text{o}}$ & $\wh{\tau}_{\text{ig}}^{\text{ps}}$ & $\wh{\tau}_{\text{ig}}^{\text{o}}$ & $\wh{\tau}^{\text{ps}}$ & $\wh{\tau}^{\text{o}}$ & $\wh{\tau}^{\text{dr}}$ &  $\wh{\tau}_{\mc{A}}^{\text{ps}}$ & $\wh{\tau}_{\mc{A}}^{\text{dr}}$ \\[1.5ex]
         \hline &&&&&&&&&&&&&\\[-1ex]
Est. & 2.54     & 3.42   & 3.46   & 3.45 & 3.55 & 3.52 & {\bfseries 3.41} & 3.32 & {\bfseries 3.45} & 3.36 & {\bfseries 3.45} & {\bfseries 3.45} & {\bfseries 3.46} \\[1.5ex] 
SD       & .45$^{\star}$      & .48$^{\star}$   &    .44$^{\star}$   & .48$^{\star}$    & 0.20 & 0.21 & .13 & .08 & .18 & .14 & .16 & 2.92 & 1.79 \\
\hline
\end{tabular}
\vskip1ex
Legend:\\ Asterisked estimators are based on the original data, i.e., in the absence of linkage error. ``SD" (starred) refers to the associated sandwich standard deviation estimate; for the others, ``SD" refers to the randomness of the linkage error varying over 100k replications. ``Est."~are averages.\\ $^{\text{pl}}$ --- estimator based on the ``plain" mean difference in the two treatment groups,   \\
$^{\text{naive}}$ --- estimators based on the subset of correctly matched data, \\
$^{\text{ig}}$ --- estimators based on using the full data, ignoring linkage error.\\
$^{\text{ps}}$, $^{\text{o}}$,  $^{\text{dr}}$, $\mc{A}$ --- propensity score, outcome model, and DR-type
estimators; $\mc{A}$ indicates restriction to audit sample (10\% of the observations). 
}
\caption{Comparison of different estimators of the ATE in the case study, without (first 4 columns) and with linkage error (remaining columns).}\label{tab:casestudy}
\end{table}

\section{Conclusion}\label{sec:conclusion}
We have considered the traditional causal analysis framework for 
an average treatment effect for a binary exposure in the absence of unobserved
confounders. This paper focuses on the challenges that arise when exposure,
outcome, and confounders originate in two separate files, which are combined via record linkage for the purpose of joint analysis. The underlying setup is that of secondary analysis in which only the linked file is available and information about the linkage process is limited. Specifically, we have studied the impact of mismatch error analytically and empirically, and have devised a framework for statistical inference that adjusts for such error. Our approach ensures the identifiability of the average treatment effect under suitable assumptions, and asymptotic inference can be conducted by leveraging methodology concerning estimating equations with latent variables. This paper prompts various directions of future research. One open 
question raised herein concerns multiple robustness safeguarding against 
an incorrect model for the latent mismatch indicators. Second, it is worth
studying to what extent approaches based on multiple imputation of these indicators 
(as considered in \cite{Guha2024} and \cite{Shan2021}) might simplify and generalize inference. Finally, missing links are pervasive in applications. This work has tacitly assumed that missing links are ignorable; it is desirable to extend the framework so that both missing and incorrect links can be handled simultaneously.  

\bibliographystyle{plain}
\bibliography{causal_linkage}

\begin{thebibliography}{10}

\bibitem{abowd2021}
J.~Abowd, J.~Abramowitz, M.C. Levenstein, K.~McCue, D.~Patki, T.~Raghunathan, A.M. Rodgers, M.D. Shapiro, N.~Wasi, and D.~Zinsser.
\newblock {Finding Needles in Haystacks: Multiple-Imputation Record Linkage Using Machine Learning}.
\newblock {Federal Reserve Bank of Boston Research Department Working Papers No. 22-11}, 2021.

\bibitem{LifeMcit}
M.~Bailey, P.Z. Lin, A.R.~Shaqir Mohammed, P.~Mohnen, J.~Murray, M.~Zhang, and A.~Prettyman.
\newblock {LIFE-M: The Longitudinal, Intergenerational Family Electronic Micro-Database}.
\newblock Inter-university Consortium for Political and Social Research (ICPSR), December 2022.

\bibitem{Binette2022}
O.~Binette and R.~Steorts.
\newblock {(Almost) all of entity resolution}.
\newblock {\em Science Advances}, 8(12):eabi8021, 2022.

\bibitem{Bukke2025}
P.~Bukke and M.~Slawski.
\newblock Relaxing the assumption of strongly non-informative linkage error in secondary regression analysis of linked files.
\newblock arXiv:2510.17553.

\bibitem{Chambers2009}
R.~Chambers.
\newblock Regression analysis of probability-linked data.
\newblock Technical report, Statistics New Zealand, 2009.

\bibitem{Chambers2023}
R.~Chambers, E.~Fabrizi, M.~Ranalli, N.~Salvati, and S.~Wang.
\newblock Robust regression using probabilistically linked data.
\newblock {\em WIREs Computational Statistics}, 15(2):e1596, 2023.

\bibitem{Christen2012}
P.~Christen.
\newblock {\em Data Matching: Concepts and Techniques for Record Linkage, Entity Resolution, and Duplicate Detection}.
\newblock Springer, 2012.

\bibitem{Dempster1977}
A.~Dempster, N.~Laird, and D.B. Rubin.
\newblock Maximum likelihood from incomplete data via the em algorithm.
\newblock {\em Journal of the Royal Statistical Society Series B}, 39(1):1--22, 1977.

\bibitem{DataFirst}
G.~Eaton, K.~Hill, and A.~Summerfield.
\newblock Data first: Criminal courts linked data research report.
\newblock {\em International Journal of Population Data Science}, 7(3):1920, 2022.

\bibitem{Elashoff2004}
M.~Elashoff and L.~Ryan.
\newblock {An EM algorithm for estimating equations}.
\newblock {\em Journal of Computational and Graphical Statistics}, 13:48--65, 2004.

\bibitem{Enamorado2019}
T.~Enamorado, B.~Fifield, and K.~Imai.
\newblock Using a probabilistic model to assist merging of large-scale administrative records.
\newblock {\em American Political Science Review}, 113:353--371, 2019.

\bibitem{N3C}
K.~Gersing.
\newblock {National COVID-19 Longitudinal Research Database Linked to Medicare and Medicaid Data}.
\newblock Technical report, U.S.~Department of Health \& Human Services, National Center for Advancing Translational Sciences, 2024.

\bibitem{Guha2024}
S.~Guha and J.~Reiter.
\newblock Regression-assisted bayesian record linkage for causal inference in observational studies with covariates spread over two files.
\newblock {\em Journal of Statistical Planning and Inference}, 229:106090, 2024.

\bibitem{Guha2022}
S.~Guha, J.~Reiter, and A.~Mercatanti.
\newblock Bayesian causal inference with bipartite record linkage.
\newblock {\em Bayesian Analysis}, 17(4):1275--1299, 2022.

\bibitem{gutmanmixture}
R.~Gutman, C.J. Sammartino, T.C. Green, and B.T. Montague.
\newblock Error adjustments for file linking methods using encrypted unique client identifier (euci) with application to recently released prisoners who are hiv+.
\newblock {\em Statistics in Medicine}, 35(1):115--129, 2016.

\bibitem{Hernan2010book}
M.~Hernan and J.~Robins.
\newblock {\em Causal Inference: What If}.
\newblock CRC Press, 2010.

\bibitem{Hernan2010}
M.A. Hernan and J.M. Robins.
\newblock {\em Causal Inference}.
\newblock CRC Boca Raton, FL, 2010.

\bibitem{Horn2012}
R.A. Horn and C.R. Johnson.
\newblock {\em Matrix Analysis}.
\newblock Cambridge University Press, 2012.

\bibitem{Imbens2015}
G.~Imbens and D.~Rubin.
\newblock {\em Causal Inference in Statistics, Social, and Biomedical Sciences}.
\newblock Cambridge University Press, 2015.

\bibitem{Kamat2024}
G.~Kamat and R.~Gutman.
\newblock Analysis of linked files: A missing data perspective.
\newblock {\em Statistical Science}, forthcoming, 2024.

\bibitem{Kang2007}
J.~Kang and J.~Schafer.
\newblock Demystifying double robustness: A comparison of alternative strategies for estimating a population mean from incomplete data.
\newblock {\em Statistical Science}, 22(4):523--539, 2007.

\bibitem{Lahiri05}
P.~Lahiri and M.~D. Larsen.
\newblock Regression analysis with linked data.
\newblock {\em Journal of the American Statistical Association}, 100(469):222--230, 2005.

\bibitem{Lunceford2004}
J.~Lunceford and M.~Davidian.
\newblock Stratification and weighting via the propensity score in estimation of causal treatment effects: a comparative study.
\newblock {\em Statistics in Medicine}, 23(19):2937--2960, 2004.

\bibitem{Hernan2025}
\url{https://miguelhernan.org/whatifbook}. Retrieved 10/15/2025.

\bibitem{Neter65}
J.~Neter, S.~Maynes, and R.~Ramanathan.
\newblock The effect of mismatching on the measurement of response error.
\newblock {\em Journal of the American Statistical Association}, 60:1005--1027, 1965.

\bibitem{Newcombe}
H.~Newcombe and J.~Kennedy.
\newblock Record linkage: making maximum use of the discriminating power of identifying information.
\newblock {\em Communications of the ACM}, 5(11):563--566, 1962.

\bibitem{Pananjady2017}
A.~Pananjady, M.~Wainwright, and T.~Courtade.
\newblock Linear regression with shuffled data: Statistical and computational limits of permutation recovery.
\newblock {\em IEEE Transactions on Information Theory}, 64(5):3286--3300, 2017.

\bibitem{Robins1994}
J.~Robins, A.~Rotnitzky, and L.~Zhao.
\newblock Estimation of regression coefficients when some regressors are not always observed.
\newblock {\em Journal of the American Statistical Association}, 89(427):846--866, 1994.

\bibitem{Rosenbaum2010}
P.~Rosenbaum.
\newblock {\em Design of Observational Studies}.
\newblock Springer, 2020.

\bibitem{Rosenbaum1983}
P.~Rosenbaum and D.~Rubin.
\newblock The central role of the propensity score in observational studies for causal effects.
\newblock {\em Biometrika}, 70(1):41--55, 1983.

\bibitem{Shan2021}
M.~Shan, K.~Thomas, and R.~Gutman.
\newblock A multiple imputation procedure for record linkage and causal inference to estimate the effects of home-delivered meals.
\newblock {\em The Annals of Applied Statistics}, 15(1):412, 2021.

\bibitem{Slawski2019}
M.~Slawski and E.~Ben-David.
\newblock Linear regression with sparsely permuted data.
\newblock {\em Electronic Journal of Statistics}, 13:1--36, 2019.

\bibitem{Slawski2021}
M.~Slawski, G.~Diao, and E.~Ben-David.
\newblock A pseudo-likelihood approach to linear regression with partially shuffled data.
\newblock {\em Journal of Computational and Graphical Statistics}, 30(4):991--1003, 2021.

\bibitem{Slawski2024}
M.~Slawski, B.T. West, P.~Bukke, Z.~Wang, G.~Diao, and E.~Ben-David.
\newblock A general framework for regression with mismatched data based on mixture modelling.
\newblock {\em Journal of the Royal Statistical Society Series A: Statistics in Society}, 188(3):896--919, 2025.

\bibitem{Stefanski2002}
L.~Stefanski and D.~Boos.
\newblock {The calculus of M-estimation}.
\newblock {\em The American Statistician}, 56(1):29--38, 2002.

\bibitem{Wang2022}
Z.~Wang, E.~Ben-David, G.~Diao, and M.~Slawski.
\newblock Regression with linked datasets subject to linkage error.
\newblock {\em WIREs Computational Statistics}, 14(4):e1570, 2022.

\end{thebibliography}
\vskip8ex

\noindent {\Large {\bfseries \textsf{Appendix}}}

\renewcommand{\thesubsection}{\Alph{subsection}}
\subsection{Proof of Proposition \ref{prop:bias_naive}}
We start with Scenario I. Let $Y^*$ denote the outcome obtained through linkage with File A, and let
$Y'$ be identically distributed as $Y$.  
\begin{align*}
\E[Y^* I(E = 1)/p_{\bm{\phi}}(X)] &= \E[Y \, I(E = 1)/p_{\bm{\phi}}(X) \, I(M = 0)] + \E[Y' \, I(E = 1)/p_{\bm{\phi}}(X) \, I(M = 1)]  \\
&= (1-\alpha) \E \E[Y^{(1)} \, I(E = 1)/p_{\bm{\phi}}(X) | X] + \alpha \E[Y'] \E[I(E = 1)/p_{\bm{\phi}}(X)] \\
&= (1-\alpha) \E[Y^{(1)}] + \alpha \E[Y],
\end{align*}
where we have used that $M$ is independent of all other variables and that for mismatched observations ($M = 1$), the outcome $Y'$ (in File B) is independent of $(X,E)$ (in File A) by ({\bfseries A5}). Likewise, one obtains $\E[Y^* I(E = 0)/p_{\bm{\phi}}(X)] = (1-\alpha) \E[Y^{(0)}] + \alpha \E[Y]$ and the result follows.  

Turning to Scenario II, let $X^*$ denote the covariates obtained through linkage with File A. Let further $X'$ be identically distributed as $X$. We have 
\begin{align*}
\E[Y I(E = 1)/p_{\bm{\phi}}(X^*)] &= \E[Y^{(1)} I(E = 1)/p_{\bm{\phi}}(X) \, I(M = 0)] +  \E[Y^{(1)} I(E = 1)/p_{\bm{\phi}}(X') \, I(M = 1)]   \\
&=(1-\alpha) \E[Y^{(1)}] + \alpha \E\left[\frac{Y^{(1)}  I(E=1)}{p_{\bm{\phi}}(X)} \frac{p_{\bm{\phi}}(X)}{p_{\bm{\phi}}(X')}\right] \\
&=(1-\alpha) \E[Y^{(1)}] + \alpha \E_{X,X'} \E\left[\frac{Y^{(1)}  I(E=1)}{p_{\bm{\phi}}(X)} \frac{p_{\bm{\phi}}(X)}{p_{\bm{\phi}}(X')} \, \Big | X, X'\right]  \\
&= (1-\alpha) \E[Y^{(1)}] + \alpha \E_{X,X'} \left[ \E[Y^{(1)} | X, X'] \E \left[\frac{I(E = 1)]}{p_{\bm{\phi}}(X)} \frac{p_{\bm{\phi}}(X)}{p_{\bm{\phi}}(X')}  \, \Big | X, X' \right] \right] \\
&= (1-\alpha) \E[Y^{(1)}] + \alpha \E_{X,X'} \left[ \E[Y^{(1)} | X] \frac{p_{\bm{\phi}}(X)}{p_{\bm{\phi}}(X')} \right] \\
&= (1-\alpha) \E[Y^{(1)}] + \alpha \E_{X,X'} \left[ \mu_1(X) \frac{p_{\bm{\phi}}(X)}{p_{\bm{\phi}}(X')} \right],
\end{align*}
where we have used that $M$ is independent of all other variables and that for a mismatch record $(X', E, Y)$,  we have $X' \indep (X,E,Y)$ by ({\bfseries A5}). The expectation $\E[Y I(E = 0) / (1 - p_{\bm{\phi}}(X^*))]$ can be evaluated analogously.

Lastly, we study Scenario III. let $E^*$ denote the exposure obtained through linkage with File $B$. We have 
\begin{align*}
\E[Y I(E^* = 1)/p_{\bm{\phi}}(X)] &= \E[Y I(E^* = 1)/p_{\bm{\phi}}(X) \, I(M = 0)] +  \E[Y I(E^* = 1)/p_{\bm{\phi}}(X) \, I(M = 1)] \\
&= (1-\alpha) \E[Y^{(1)}] + \alpha \big \{ \E[Y I(E^* = 1)/p_{\bm{\phi}}(X) \, I(E = 1)] +\\
&\qquad \qquad \qquad \qquad \quad \quad + \E[Y I(E^* = 1)/p_{\bm{\phi}}(X) \, I(E = 0)] \big \} \\
&= (1-\alpha) \E[Y^{(1)}] + \alpha p \E[Y^{(1)}] + \alpha p \E_X \left[\mu_0(X) \frac{1 - p_{\bm{\phi}}(X)}{p_{\bm{\phi}}(X) } \right],
\end{align*}
where we have used that $M$ is independent of all other variables, the fact that $E^*$ originating from a different record ($M = 1$) is independent of $(Y,E,X)$ by ({\bfseries A5}), and 
\begin{align*}
\E[Y I(E = 0) / p_{\bm{\phi}}(X)] &= \E_X \E[Y^{(0)} I(E = 0) / p_{\bm{\phi}}(X) | X] \\
                                  &= \E_X \left[ \E[Y^{(0)}|X] \E[ I(E = 0) / p_{\bm{\phi}}(X) | X] \right] \\ & = \E_X \left[ \mu_0(X) \left[ \frac{ 1 - p_{\bm{\phi}}(X)}{p_{\bm{\phi}}(X)} \right] \right],
\end{align*}
with $\mu_0(X) = \E[Y^{(0)} | X]$. The expectation $\E[Y I(E^* = 0) / \{ 1 - p_{\bm{\phi}}(X) \}]$ can be evaluated analogously. \qed

\subsection{Proof of Proposition \ref{prop:unbiased_pso}}
We have 
\begin{align*}
\E \left[\frac{Y I(E = 1) I(M = 0)}{(1-h_{\bm{\gamma}}(Z)) p_{\bm{\phi}}(X)} \right]
&= \E \left[ \E \left[\frac{Y I(E = 1) I(M = 0)}{(1-h_{\bm{\gamma}}(Z)) p_{\bm{\phi}}(X)} \Big | X, Z \right] \right] \\
&\overset{ \text{{\bfseries (A1), (A2)}}}{=}  
\E \left[ \E \left[\frac{Y I(E = 1)}{p_{\bm{\phi}}(X)} \Big | X, Z \right] \E \left[\frac{I(M=  0)}{1-h_{\bm{\gamma}}(Z)} \Big| Z \right]\right] \\
&\overset{ \text{{\bfseries (A3)}}}{=} \E \left[ \E \left[\frac{Y I(E = 1)}{p_{\bm{\phi}}(X)} \Big | X\right] \right]  \\
&\overset{ \text{{\bfseries (A1)}}}{=} \E [\E[Y^{(1)}|X]] = \E[Y^{(1)}]. 
\end{align*}
An analogous argument can be made when replacing $I(E = 1)$ by $I(E = 0)$ and 
$p_{\bm{\phi}}$ by $1 - p_{\bm{\phi}}$, respectively. \qed

\subsection{Derivation of \eqref{eq:marginal_scenario1}}\label{app:marginal_scenario1}
Let $P$ and $P_Z$ be the probability measures with mass $1/n$ at each of the triplets 
$\{ (\M{x}_i, e_i, \M{z}_i) \}_{i = 1}^n$ and each of the $\{ \M{z}_i \}_{i = 1}^n$, respectively, and let $f_{Y|M=1, X=\M{x}, E=e, Z = \M{z}}$ denote the conditional density of $Y$ given the quantities after the dash $|$ in the subscript. We have 
{\small \begin{align*}
f_{Y|M=1}(y) &\overset{{(i)}}{=} \int f_{Y|M=1, X = \M{x}, E = e, Z = \M{z}}(y) \; \frac{\p(M = 1 | X = \M{x}, E = e, Z = \M{z})}{\int \p(M = 1 | X = \M{x}', E = e', Z = \M{z}') \; dP(\M{x}', e', \M{z'})}  \;dP(\M{x}, e, \M{z})  \\
&\overset{{(ii)}}{=} \int f_{Y|X = \M{x}, E = e}(y) \; \frac{\p(M = 1 | Z = \M{z})}{\int \p(M = 1 | Z = \M{z}') \, dP_{Z}(\M{z'})}  \;dP(\M{x}, e, \M{z}) \\
&\overset{(iii)}{=} \int  \varphi_{\sigma}(y - \mu_{\bm{\beta}}^{e}(\M{x})) \; \frac{\p(M = 1 | Z = \M{z})}{\int \p(M = 1 | Z = \M{z}') \, dP_{Z}(\M{z'})}  \;dP(\M{x}, e, \M{z}) \\
&= \int  \varphi_{\sigma}(y - \mu_{\bm{\beta}}^{e}(\M{x})) \; \frac{\p(M = 1 | Z = \M{z})}{\frac{1}{n} \sum_{j = 1}^n \p(M = 1 | Z = \M{z}_j)}  \;dP(\M{x}, e, \M{z}) \\
&= \frac{1}{n} \sum_{i = 1}^n \varphi_{\sigma}(y - \mu_{\bm{\beta}}^{e_i}(\M{x}_i)) \; \frac{\p(M = 1 | Z = \M{z}_i)}{\frac{1}{n} \sum_{j = 1}^n \p(M = 1 | Z = \M{z}_j)} \\
&\overset{(iv)}{=} \sum_{i = 1}^n \varphi_{\sigma}(y - \mu_{\bm{\beta}}^{e_i}(\M{x}_i)) \frac{h_{\bm{\gamma}}(\M{z}_i)}{\sum_{j = 1}^n h_{\bm{\gamma}}(\M{z}_j)} =  \sum_{i = 1}^n \varphi_{\sigma}(y - \mu_{\bm{\beta}}^{e_i}(\M{x}_i)) \, w(\M{z}_i),  
\end{align*}}
where in (i) we have used Bayes' formula, and in (ii) we have invoked Assumptions {\bfseries (A1)}, {\bfseries (A2)}, and {\bfseries (A3)}. In (iv) and (v) we invoke the specific models \eqref{eq:outcome_model} and \eqref{eq:prop_logistic}, respectively. \qed     

\subsection{Derivation of \eqref{eq:marginal_scenario2}}\label{app:marginal_scenario2}
With similar arguments as in the preceding subsection, we obtain the following. 
{\small \begin{align*}
&f_{Y|E = e, M=1}(y) \cdot \p(E = e | M = 1)  \\
&= \int
f_{Y|E = e, X = \M{x}, Z = \M{z}, M=1}(y) \cdot \p(E = e | X = \M{x}, Z = \M{z}, M = 1) \cdot \frac{\p(M = 1 | Z = \M{z})}{\int \p(M = 1 | Z = \M{z}') \, dP_{Z}(\M{z'})}  \;dP(\M{x},\M{z}) \\
&= \int
f_{Y|E = e, X = \M{x}}(y) \cdot \p(E = e | X = \M{x}) \cdot \frac{\p(M = 1 | Z = \M{z})}{\int \p(M = 1 | Z = \M{z}') \, dP_{Z}(\M{z'})}  \;dP(\M{x},\M{z}) \\
&=  \int  \varphi_{\sigma}(y - \mu_{\bm{\beta}}^{e}(\M{x}))  \cdot \{ p_{\bm{\phi}}(\M{x})^e \cdot (1 - p_{\bm{\phi}}(\M{x}))^{1-e} \} \cdot \frac{\p(M = 1 | Z = \M{z})}{\int \p(M = 1 | Z = \M{z}') \, dP_{Z}(\M{z'})}  \;dP(\M{x},\M{z}) \\
&= \su [ \varphi_{\sigma}(y - \mu_{\bm{\beta}}^{e}(\M{x}_i))  \{ p_{\bm{\phi}}(\M{x}_i)^{e} (1 - p_{\bm{\phi}}(\M{x}_i))^{1 - e}  \} \cdot w(\M{z}_i)].
\end{align*}}

\subsection{Derivation of \eqref{eq:marginal_scenario3}}\label{app:marginal_scenario_3}
The first expression follows from the fact that 
\begin{align*}
f_{Y|X = \M{x}_i, M = 1}(y) &= \sum_{e = 0}^1 f_{Y|X = \M{x}_i, E = e, M = 1}(y) \, \p(E = e|X = \M{x}_i, M = 1) \\
&= \sum_{e = 0}^1 f_{Y|X = \M{x}_i, E = e}(y) \, \p(E = e|X = \M{x}_i) \\
&= \varphi_{\sigma}(y - \mu_{\bm{\beta
}}^1) p_{\bm{\phi}}(\M{x}_i) + \varphi_{\sigma}(y - \mu_{\bm{\beta
}}^0) (1 - p_{\bm{\phi}}(\M{x}_i)), 
\end{align*}
where the second identity uses assumptions {\bfseries (A1)} and {\bfseries (A2)}. 
The derivation of the expression for $\p(E = e | M = 1)$ is completely along the lines of the arguments made in the preceding subsections, and is thus omitted. 

\subsection{Derivation of \eqref{eq:marginal_scenario2_audit}}\label{app:marginal_extra}
For $e \in \{0,1\}$ and $y \in \R$, we have 
\begin{align*}
f_{Y|E = e, M = 1}(y) &= \int f_{Y|X = \M{x}, Z = \M{z}, E=e, M = 1}(y) \; \frac{\p(M = 1, E = e | \M{x}, \M{z})}{\int \p(M = 1, E = e | \M{x}', \M{z}') \; dP(\M{x}', \M{z}')} \;\,dP(\M{x}, \M{z}) \\
&=  \int f_{Y|X = \M{x}, E=e}(y) \; \frac{\p(M = 1, E = e | \M{x}, \M{z})}{\int \p(M = 1, E = e | \M{x}', \M{z}') \; dP(\M{x}', \M{z}')} \;\,dP(\M{x}, \M{z}) \\
&= \int f_{Y|X = \M{x}, E=e}(y) \; \frac{\p(M = 1| Z = \M{z}) \cdot \p(E = e | X = \M{x})}{\int \p(M = 1| Z = \M{z}') \cdot  \p(E = e| X = \M{x}') \; dP(\M{x}', \M{z}')} \;\,dP(\M{x}, \M{z}) \\
&= \int \varphi_{\sigma}(y - \mu_*(\M{x}, e)) \; \frac{\p(M = 1| Z = \M{z}) \cdot \p(E = e | X = \M{x})}{\int \p(M = 1| Z = \M{z}') \cdot  \p(E = e| X = \M{x}') \; dP(\M{x}', \M{z}')} \;\,dP(\M{x}, \M{z})  \\
&=\su   w_{e}(\M{z}_i, \M{x}_i) \varphi_{\sigma}(y - \mu_*(\M{x}_i, e)), 
\end{align*}
where 
\begin{equation*}
w_e(\M{z}_i, \M{x}_i) = \frac{h_{\bm{\gamma}}(\M{z}_i) \cdot \{ p_{\bm{\phi}}(\M{x}_i) \}^{e} \{ 1 - p_{\bm{\phi}}(\M{x}_i) \}^{1 - e}}{\sum_{k = 1}^n [h_{\bm{\gamma}}(\M{z}_k) \cdot \{ p_{\bm{\phi}}(\M{x}_k) \}^{e} \{ 1 - p_{\bm{\phi}}(\M{x}_k) \}^{1 - e}] }, \quad 1 \leq i \leq n.
\end{equation*}

\subsection{Proof of Proposition \ref{prop:unbiased_est}}
There is nothing to show when $\lambda_2 = 0$. Let us next assume $\lambda_2 = 1$ and $\lambda_3 = 0$. For what 
follows, we drop the observation index $i$ and instead consider the random variables 
$(X,Y,Z, E, M, \wh{M})$ underlying the corresponding quantities denoted by lowercase letters. We have 
\begin{align*}
&\E_{\bm{\beta}, \bm{\gamma}, \bm{\phi}} \left[  \{ 1- \wh{M}(\bm{\beta}, \bm{\gamma}) \}  \frac{Y I(E = 1)}
{(1 - h_{\bm{\gamma}}(Z)) p_{\bm{\phi}}(X)}  \right] \\
&= \E_{\bm{\beta}, \bm{\gamma}, \bm{\phi}} \left[  \E\left[ \{ 1- M \}  | X, Y, Z \right] \frac{Y I(E = 1)}
{(1 - h_{\bm{\gamma}}(Z)) p_{\bm{\phi}}(X)}  \right] \\
&\overset{ \text{{\bfseries (A1), (A2)}}}{=} \E_{\bm{\beta}, \bm{\gamma}, \bm{\phi}} \left[  \E\left[ \{ 1- M \} \frac{Y I(E = 1)}
{(1 - h_{\bm{\gamma}}(Z)) p_{\bm{\phi}}(X)}  \Big| X, Y, Z \right]   \right] \\
&= \E_{\bm{\beta}, \bm{\gamma}, \bm{\phi}} \left[ \frac{I(M= 0)I(E = 1) Y} 
{(1 - h_{\bm{\gamma}}(Z)) p_{\bm{\phi}}(X)}   \right] = \E[Y^{(1)}],
\end{align*}
where the last equality is obtained by following the steps in the proof of Proposition \ref{prop:unbiased_pso}. Replacing 
$I(E = 1)$ by $I(E = 0)$ and $p_{\bm{\phi}}$ by $1 - p_{\bm{\phi}}$, we similarly obtain $\E[Y^{(0)}]$, so that 
$\E[Y^{(1)}] - \E[Y^{(0)}] - \tau = 0$, as needed to be shown. 

Finally, consider $\lambda_1 = \lambda_2 = \lambda_3 = 1$. With an argument parallel to the one used in 
the previous display, we obtain that
\begin{align*}
&\E_{\bm{\beta}, \bm{\gamma}, \bm{\phi}} \left[  \{ 1- \wh{M}(\bm{\beta}, \bm{\gamma}) \}  \frac{(Y - \mu_{\bm{\beta}}^1(X)) I(E = 1)}
{(1 - h_{\bm{\gamma}}(Z)) p_{\bm{\phi}}(X)}  \right] \\
&= \E_{\bm{\beta}, \bm{\gamma}, \bm{\phi}} \left[    \frac{(Y - \mu_{\bm{\beta}}^1(X)) I(E = 1) I(M = 0)}
{(1 - h_{\bm{\gamma}}(Z)) p_{\bm{\phi}}(X)}  \right] \\
&= \E_{\bm{\beta}, \bm{\gamma}, \bm{\phi}} \E \left[    \frac{(Y - \mu_{\bm{\beta}}^1(X)) I(E = 1) I(M = 0)}
{(1 - h_{\bm{\gamma}}(Z)) p_{\bm{\phi}}(X)}  \Big| X,Z \right] = \E \E_{\bm{\beta}}[(Y^{(1)} - \mu_{\bm{\beta}}^1(X))|X] = 0.
\end{align*}
Using a parallel argument for $E = 0$, we achieve a reduction to the case $\lambda_2 = 0$, which is trivial as noted above. \qed

\subsection{Efficient computation of the covariance matrix \eqref{eq:covariance_est}}\label{app:eff_cov}
Consider the Jacobian
\begin{equation*}
J (\bm{\theta}, \M{m}) = \begin{pmatrix}
                        \frac{\partial}{\partial \bm{\theta}} Q_{\bm{\theta}}(\bm{\theta}, \M{m}) & \frac{\partial}{\partial \M{m}} Q_{\bm{\theta}}(\bm{\theta}, \M{m}) \\[2ex]
                        \frac{\partial}{\partial \bm{\theta}} Q_{\M{m}}(\bm{\theta}, \M{m}) &  \frac{\partial}{\partial \M{m}} Q_{\M{m}}(\bm{\theta}, \M{m})
                        \end{pmatrix}  = \begin{pmatrix}
                        A  & B  \\[2ex]
                        C &  - I_n
                        \end{pmatrix},  
\end{equation*}                        
where $A$, $B$ and $C$ are shortcuts from the top diagonal block and the two off-diagonal blocks, respectively, and we have used that by construction of $Q_{\M{m}}(\bm{\theta}, \M{m})$, we have 
$\frac{\partial}{\partial \M{m}} Q_{\M{m}}(\bm{\theta}, \M{m}) = -I_n$. Using the partitioned
inverse formula based on Schur complements \cite[][$\S$0.7.3.]{Horn2012}, we have 
\begin{align*}
\{ J (\bm{\theta}, \M{m}) \}^{-1} = \begin{pmatrix}
S^{-1} & -S^{-1} B (-I_n)  \\[1ex]
-(-I_n) C S^{-1} & -I_n - I_n C S^{-1} B (-I_n)
\end{pmatrix}
\end{align*}
where $S = A - B (-I_n)  C$ is the Schur complement of $J(\bm{\theta}, \M{m}
)$ w.r.t.~the bottom diagonal block. Observe that $[ \{ J(\bm{\theta}, \M{m}) \}^{-1}]_{\bm{\theta} \bm{\theta}} = S^{-1}$. Computing $S$ and $S^{-1}$ requires $O(d^2 n + d^3)$ flops. 

\end{document}